\documentstyle{article} 
\newtheorem{theorem}{Theorem}[section]

\newtheorem{ex}{Example}
\newtheorem{definition}[theorem]{Definition}

\newcommand{\auth}{\mbox{auth}}

\makeatletter
\let\@xpar=\par
\newenvironment{proof}{{\vspace{1ex}\@xpar\noindent}%
{\bf Proof.}\hspace{3mm} }{\penalty10000\penalty10000%
\hfill$\Box$\vspace{.5ex}\@xpar\penalty-8000\noindent}

\def\setcounterrefvalue#1#2{\@ifundefined{c@#1}{\@nocnterr}%
{\@ifundefined{r@#2}{\edef\@tempb{\z@}}{\edef\@tempc{\@nameuse{r@#2}}
       \relax\edef\@tempb{\expandafter\@car\@tempc\@nil}}
\global\csname c@#1\endcsname \@tempb\relax}}

\newcount\@algordepth \@algordepth = 0

\@definecounter{algori}
\@definecounter{algorii}

\def\setcounterrefvalue#1#2{\@ifundefined{c@#1}{\@nocnterr}%
\@ifundefined{r@#2}{\edef\@tempb{\z@}}{\edef\@tempc{\@nameuse{r@#2}}
       \relax\edef\@tempb{\expandafter\@car\@tempc\@nil}
\global\csname c@#1\endcsname \@tempb\relax}}

\def\algorithm#1{\ifnum \@algordepth >1 \@toodeep\else
      \advance\@algordepth \@ne
      \edef\@algorctr{algor\romannumeral\the\@algordepth}\list
      {\hspace\labelsep \csname label\@algorctr\endcsname}%
{\settowidth\labelwidth{\hspace\labelsep {\bf Step #1}}%
\leftmargin\labelwidth
\advance\leftmargin\labelsep\usecounter{\@algorctr}}
\setlength{\itemsep}{2.0pt}
\setlength{\topsep}{4.5pt plus 1pt minus 1pt}%
\fi}

\def\thealgori{\arabic{algori}}

\def\p@algorii{\thealgori.}

\def\p@enumii{\theenumi.}

\begin{document}
\bibliographystyle{plain}

\title{Perfectly Secure Message Transmission Revisited\thanks{An extended
abstract of this paper has appeared in \cite{dwe02}}}

\author{Yvo Desmedt\\
Computer Science, Florida State University, Tallahassee\\
Florida FL 32306-4530, USA, {\tt desmedt@cs.fsu.edu}\\
\and
Yongge Wang\\
Department of Software and Information Systems,\\
University of North Carolina at Charlotte,\\
9201 University City Blvd, Charlotte, NC 28223,\\
{\tt yonwang@uncc.edu}
}

\maketitle
\begin{abstract}
Achieving secure communications in networks has been one
of the most important problems in information technology.
Dolev, Dwork, Waarts, and Yung have studied secure
message transmission in one-way or two-way channels.
They only consider the case when all channels are two-way or
all channels are one-way. Goldreich, Goldwasser, and Linial,
Franklin and Yung, Franklin and Wright, and Wang and
Desmedt have studied secure communication and
secure computation in multi-recipient (multicast)
models. In a ``multicast channel'' (such as Ethernet), one processor
can send the same message---simultaneously and privately---to
a fixed subset of processors. In this paper, we shall
study necessary and sufficient conditions for achieving
secure communications against active adversaries in mixed
one-way and two-way channels. We also discuss
multicast channels and neighbor network channels.

\end{abstract}
\section{Introduction}
If there is a private and authenticated channel between two parties,
then secure communication between them is guaranteed.
However, in most cases, many parties are only indirectly
connected, as elements of an incomplete network of private
and authenticated channels.
In other words they need to use intermediate or internal nodes.
Achieving participants cooperation in the presence of faults
is a major problem in distributed networks.
Original work on secure distributed computation assumed a
complete graph for secure and reliable communication.
Dolev, Dwork, Waarts, and Yung \cite{ddwy} were able
to reduce the size of the network graph by providing
protocols that achieve private and reliable communication
without the need for the parties to start with secret keys.
The interplay of network connectivity and secure
communication has been studied extensively
(see, e.g., \cite{bgw,ccd,dolev,ddwy,ha}).
For example, Dolev \cite{dolev} and Dolev et al. \cite{ddwy} showed
that, in the case of $k$ Byzantine faults, reliable communication
is achievable only if the system's network is
$2k+1$ connected. They also showed that if all the paths are one
way, then $3k+1$ connectivity is necessary and sufficient
for reliable and private communications.
However they did not prove any results for the general case
when there are certain number of directed paths in one direction
and another number of directed paths in the other direction.
While undirected graphs correspond naturally to the case of pairwise
two-way channels, directed graphs do not correspond to the case of
all-one-way or all-two-way channels considered in \cite{ddwy},
but to the mixed case where there are some paths in one direction
and some paths in the other direction.
In this paper, we will initiate the study in this direction
by showing what can be done with a general directed graph.
Note that this scenario is important in practice, in particular,
when the network is not symmetric. For example, a channel from $A$ to
$B$ is cheap and a channel from $B$ to $A$ is expensive but not impossible.
Another example is that $A$ has access to more resources than $B$ does.

Goldreich, Goldwasser, and Linial \cite{ggl},
Franklin and Yung \cite{fy}, Franklin and Wright \cite{fw},
and Wang and Desmedt \cite{wd}
have studied secure communication and
secure computation in {\it multi-recipient (multicast)}
models. In a ``multicast channel'' (such as Ethernet), one participant
can send the same message---simultaneously and privately---to
a fixed subset of participants.
Franklin and Yung \cite{fy} have given
a necessary and sufficient condition for
individuals to exchange private messages
in multicast models in the presence
of passive adversaries (passive gossipers).
For the case of active Byzantine adversaries,
many results have been presented by
Franklin and Wright \cite{fw}, and, Wang and Desmedt \cite{wd}.
Note that Goldreich, Goldwasser, and Linial \cite{ggl}
have also studied fault-tolerant computation
in the public multicast model (which can be thought of
as the largest possible multirecipient channels) in the presence
of active Byzantine adversaries. Specifically,
Goldreich, et al. \cite{ggl}
have made an investigation of general fault-tolerant distributed
computation in the full-information model.
In the full information model no restrictions are made on
the computational power of the faulty parties or
the information available to them. (Namely, the faulty
players may be infinitely powerful and there are no private
channels connecting pairs of honest players).
In particular, they present efficient two-party protocols for
fault-tolerant computation of any bivariate function.

There are many examples of multicast channels (see, e.g. \cite{fw}),
such as an Ethernet bus
or a token ring. Another example is a shared cryptographic
key. By publishing an encrypted message, a participant
initiates a multicast to the subset of  participants
that is able to decrypt it.

We present our model in Section \ref{modelsec}.
In Sections \ref{digraphsecd} and \ref{digraphsec},
we study secure message transmission over directed graphs.
Section \ref{hypersec} is devoted to reliable message transmission
over hypergraphs, and Section \ref{neighborsec} is devoted to
secure message transmission over neighbor networks.

\section{Model}
\label{modelsec}
We will abstract away the concrete network structures and
consider directed graphs. A directed graph is a graph $G(V,E)$
where all edges have directions. For a directed graph $G(V,E)$
and two nodes $A,B\in V$,
throughout this paper, $n$ denotes the number of
vertex disjoint paths between the two nodes and
$k$ denotes the number of faults under the
control of the adversary. We write $|S|$ to denote
the number of elements in the set $S$.
We write $x\in_R S$ to indicate that $x$ is chosen
with respect to the uniform distribution on $S$.
Let {\bf F} be a finite field, and let $a,b,c,M\in {\bf F}$.
We define $\auth(M;a,b) :=aM+b$ (following \cite{fw,gms,rab,rb})
and $\auth(M;a,b,c) :=aM^2+bM+c$ (following \cite{wd}).
Note that each authentication key $key=(a,b)$ can be used to
authenticate one message $M$ without
revealing any information about any component of the authentication key
and that each authentication key $key=(a,b,c)$ can be used to
authenticate two messages $M_1$ and $M_2$ without
revealing any information about any component of the authentication key.
We will also use a function $\langle \ldots \rangle$ which
maps a variable size (we assume that this variable size is 
bounded by a pre-given bound) ordered subset of {\bf F} to an image element
in a field extension ${\bf F}^*$ of {\bf F}, and from any image element 
we can uniquely and efficiently recover the ordered subset.
Let $k$ and $n$ be two integers such
that $0\leq k < n\le 3k+1$. A $(k+1)$-out-of-$n$ secret
sharing scheme is a probabilistic function
S: ${\bf F}\rightarrow {\bf F}^n$ with the property that
for any $M\in{\bf F}$ and $\mbox{S}(M)=(v_1,\ldots,v_n)$,
no information of $M$ can be inferred from any $k$ entries
of $(v_1,\ldots,v_n)$, and $M$ can
be recovered from any $k+1$ entries of $(v_1,\ldots,v_n)$.
The set of all possible $(v_1,\ldots,v_n)$ is called a code
and its elements codewords.
We say that a $(k+1)$-out-of-$n$ secret sharing scheme
can detect $k'$ errors if given any codeword $(v_1,\ldots,v_n)$ and any
tuple $(u_1,\ldots,u_n)$ over {\bf F}
such that
$0<|\{i: u_i\not= v_i, 1\le i \le n\}|\le k'$ one can detect
that $(u_1,\ldots,u_n)$ is not a codeword. If the code is
Maximal Distance Separable, then the maximum value of errors that can be
detected is $n-k-1$~\cite{MacWilliamsSloane78}.
We say that the $(k+1)$-out-of-$n$ secret sharing scheme can correct $k'$
errors if from any $\mbox{S}(M)=(v_1,\ldots,v_n)$ and any tuple
$(u_1,\ldots,u_n)$ over {\bf F} with 
$|\{i: u_i\not= v_i, 1\le i \le n\}|\le k'$
one can recover the secret $m$.
If the code is Maximal Distance Separable, then the maximum value of errors
that allows the recovery of the vector $(v_1,\ldots,v_n)$
is $(n-k-1)/2$~\cite{MacWilliamsSloane78}.
A $(k+1)$-out-of-$n$ Maximal Distance Separable (MDS) secret
sharing scheme is a $(k+1)$-out-of-$n$ secret sharing scheme
with the property that for any $k'\le (n-k-1)/2$, one can correct $k'$
errors
and simultaneously detect $n-k-k'-1$ errors (as follows easily
by generalizing~\cite[p.~10]{MacWilliamsSloane78}).
Maximal Distance Separable (MDS) secret
sharing schemes can be constructed from any MDS codes, for example,
from Reed-Solomon code \cite{ms}.

In a message transmission protocol, the sender $A$
starts with a message $M^A$ drawn from a message
space ${\cal M}$ with respect to a certain probability
distribution. At the end of the protocol, the receiver
$B$ outputs a message $M^B$. We consider a synchronous
system in which messages are sent via multicast in rounds.
During each round of the protocol, each node
receives any messages that were multicast for it
at the end of the previous round, flips
coins and perform local computations, and then
possibly multicasts a message. We will also
assume that the message space ${\cal M}$ is a
subset of a finite field {\bf F}.

We consider two kinds of adversaries.
A passive adversary (or gossiper adversary)
is an adversary who can only observe the
traffic through $k$ internal nodes.
An active adversary (or Byzantine adversary)
is an adversary with unlimited computational
power who can control $k$ internal nodes.
That is, an active adversary will not only
listen to the traffics through the controlled
nodes, but also control the message sent by
those controlled nodes.
Both kinds of adversaries are assumed to know
the complete protocol specification, message
space, and the complete structure of the
graph. In this paper, we will not consider a dynamic
adversary who could change the nodes it controls
from round to round, instead we will only consider
static adversaries. That is,
at the start of the protocol, the
adversary chooses the $k$ faulty nodes.
An alternative interpretation is that $k$ nodes are static
collaborating adversaries.

For any execution of the protocol, let $adv$ be the
adversary's view of the entire protocol. We write
$adv(M,r)$ to denote the adversary's view when
$M^A=M$ and when the sequence of coin flips used
by the adversary is $r$.
\begin{definition}
\label{poeitr}
(see Franklin and Wright \cite{fw})
\begin{enumerate}
\item Let $\delta < \frac{1}{2}$. A message transmission protocol is
$\delta$-{\rm reliable} if, with probability at least $1-\delta$, $B$
terminates
with $M^B=M^A$. The probability is over the choices of
$M^A$ and the coin flips of all nodes.
\item A message transmission protocol is {\rm reliable}
if it is $0$-{\rm reliable}.
\item A message transmission protocol is $\varepsilon$-{\rm private}
if, for every two messages $M_0,M_1$ and every $r$,
$\sum_c|\Pr[adv(M_0,r)=c]-\Pr[adv(M_1,r)=c]|\le 2\varepsilon.$
The probabilities are taken over the coin flips of the
honest parties, and the sum is over all possible values
of the adversary's view.
\item A message transmission protocol is {\rm perfectly private}
if it is $0$-{\rm private}.
\item A message transmission protocol is {\rm $(\varepsilon,\delta)$-secure}
if it is $\varepsilon$-private and $\delta$-reliable.
\item An  $(\varepsilon,\delta)$-secure  message transmission protocol is
{\rm efficient} if its round complexity and bit complexity
are polynomial in the size of the network, $\log\frac{1}{\varepsilon}$
(if $\varepsilon>0$) and $\log\frac{1}{\delta}$
(if $\delta>0$).
\end{enumerate}
\end{definition}
For two nodes $A$ and $B$ in a directed graph such that there are
$2k+1$ node disjoint paths from $A$ to $B$, there is a straightforward
reliable message transmission from $A$ to $B$ against a $k$-active
adversary:
$A$ sends the message $m$ to $B$ via all the $2k+1$ paths, and $B$ recovers
the message $m$ by a majority vote.
\section{$(0,\delta)$-Secure message transmission in directed graphs}
\label{digraphsecd}
Our discussion in this section will be concentrated on directed
graphs.  Dolev, Dwork, Waarts, and Yung \cite{ddwy} addressed the problem
of secure message transmissions in a point-to-point
network. In particular, they showed that if all channels from
$A$ to $B$ are one-way, then $(3k+1)$-connectivity is necessary
and sufficient for (0,0)-secure message transmissions  from
$A$ to $B$  against
a $k$-active adversary. They also showed that if all channels between $A$
and $B$ are two-way, then  $(2k+1)$-connectivity is necessary
and sufficient for (0,0)-secure message transmissions between
$A$ and $B$  against a $k$-active adversary.
In this section we assume that there are only
$2k+1-u$ directed node disjoint paths from $A$ to $B$,
where $1\leq u\leq k$.  We show that $u$ directed node disjoint paths
from $B$ to $A$ are necessary and sufficient to achieve
$(0,\delta)$-secure message transmissions from $A$ to $B$  against
a $k$-active adversary.

Franklin and Wright \cite{fw} showed that even if all channels 
between $A$ and $B$ are two way, $2k+1$ channels between $A$ and $B$
are still necessary for $(1-\delta)$-reliable (assuming 
that $\delta< \frac{1}{2}\left(1-\frac{1}{|{\bf F}|}\right)$)
message transmission from $A$ to $B$ against a $k$-active adversary.

\begin{theorem}
\label{necessary-minimum}
(Frandlin and Wright \cite{fw})
Let $G(V,E)$ be a directed graph, $A,B\in V$, and there are only
$2k$ two-way node disjoint paths between $A$ and $B$ in $G$.
Then $\delta$-reliable message transmission
from $A$ to $B$ against a $k$-active adversary is impossible
for $\delta< \frac{1}{2}\left(1-\frac{1}{{|\bf F}|}\right)$.
\end{theorem}

In the following, we first show that if there is no directed path
from $B$ to $A$, then $2k+1$ directed paths from
$A$ to $B$ is necessary and sufficient for $(0,\delta)$-secure message transmission
from $A$ to $B$. 
\begin{theorem}
\label{fwsuf}
Let $G(V,E)$ be a directed graph, $A,B\in V$, and $0<\delta <\frac{1}{2}$.
If there is no directed paths from $B$ to $A$, then
the necessary and sufficient condition for $(0,\delta)$-secure
message transmission from $A$ to $B$  against a $k$-active adversary
is that there are $2k+1$ directed node disjoint paths from $A$ to $B$.
\end{theorem}
\begin{proof}
The necessity is proved in Theorem \ref{necessary-minimum}.
Let $p_1,\allowbreak \ldots,\allowbreak p_{2k+1}$ be
the $2k+1$ directed node disjoint paths
from $A$ to $B$. Let $M^A\in {\bf F}$ be the secret that $A$ wants to
send to $B$. $A$ constructs $(k+1)$-out-of-$(2k+1)$ secret shares
$(s_1^A, \ldots, s_{2k+1}^A)$ of $M^A$. The protocol proceeds from round
$1$ through round $2k+1$. In each round $1\le i\le 2k+1$, 
we have the following steps:
\begin{algorithm}{9}
\item $A$ chooses
$\{(a_{i,j}^A, b_{i,j}^A)\in_R{\bf F}^2: 1\le j\le 2k+1\}$.
\item $A$ sends $(s_i^A, \auth(s_i^A;a_{i,1}^A,b_{i,1}^A), \ldots,
\auth(s_i^A;a_{i,2k+1}^A,b_{i, 2k+1}^A))$
to $B$ via path $p_i$, and sends  $(a^A_{i,j}, b_{i,j}^A)$
to $B$ via path $p_j$ for each $1\le j\le 2k+1$.
\item $B$ receives $(s^B_i, c^B_{i,1}, \ldots, c^B_{i,2k+1})$ via path
$p_i$, and receives $(a_{i,j}^B, b_{i,j}^B)$ via path $p_j$ for each
$1\le j\le 2k+1$.
\item $B$ computes $t=|\{j: c_{i,j}^B=\auth(s_i^B; a_{i,j}^B, b_{i,j}^B)\}|$.
If $t\geq k+1$, then $B$ decides that $s_i^B$ is a valid share.
Otherwise $B$ discards $s_i^B$. 
\end{algorithm}
It is easy to check that after the round
$2k+1$, with high probability, $B$ will get at least $k+1$ valid
shares of $s^A$. Thus, with high probability, $B$ will recover the secret
$M^B=M^A$. It is straightforward that the protocol achieves perfect privacy.
Thus the above protocol is a $(0,\delta)$-secure message 
transmission protocol from $A$ to $B$ against a $k$-active adversary.
\hfill{Q.E.D.}\end{proof}
By Theorem \ref{necessary-minimum}, the necessary condition for
$(0,\delta)$-secure message transmission
from $A$ to $B$ against a $k$-active adversary is that there are at least
$k+1$ node disjoint paths from $A$ to $B$ and there are at least
$2k+1$ node disjoint paths in total from $A$ to $B$ and from $B$ to $A$.
In the following, we show that this condition is also sufficient.
We first show that the condition is sufficient
for $k=1$.
\begin{theorem}
\label{forkis1}
Let $G(V,E)$ be a directed graph, $A,B\in V$. If there are
two directed node disjoint paths $p_0$ and $p_1$ from $A$ to $B$,
and one directed path $q$ (which is node disjoint from
$p_0$ and $p_1$) from $B$ to $A$, then for any $0<\delta<\frac{1}{2}$,
there is a $(0,\delta)$-secure message transmission protocol from $A$
to $B$ against a 1-active adversary.
\end{theorem}
\begin{proof}
In the following protocol, $A$ $(0,\delta)$-securely transmits
a message $M^A\in{\bf F}$ to $B$.
\begin{algorithm}{9}
\item $A$ chooses $s_0^A\in_R{\bf F}$, $(a_0^A,b_0^A),
(a_1^A,b_1^A)\in_R{\bf F}^2$, and
let $s_1^A=M^A-s_0^A$. For each $i\in\{0,1\}$, $A$ sends
$(s_i^A, (a_i^A,b_i^A),
\auth(s_{i}^A;a_{1-i}^A,b_{1-i}^A))$
to $B$ via path $p_i$.
\item Assumes that $B$ receives $(s_i^B, (a_i^B,b_i^B), c^B_i)$ via path
$p_i$.  $B$ checks whether
$c^B_i= \auth(s_{i}^B;a_{1-i}^B,b_{1-i}^B)$ for $i=0,1$.
If both equations hold, then $B$ knows that
with high probability the adversary was either passive or not on the
paths from $A$ to $B$. $B$ can recover the secret message,
sends ``OK'' to $A$ via the path $q$, and terminates the protocol.
Otherwise, one of equations does not hold and $B$ knows that the adversary
was on one of the paths from $A$ to $B$. In this case, $B$ chooses
$(a^B,b^B)\in_R{\bf F}^2$, and sends
$((a^B,b^B), (s_0^B, (a_0^B,b_0^B), c^B_0), (s_1^B, (a_1^B,b_1^B), c^B_1))$
to $A$ via the path $q$.
\item If $A$ receives ``OK'', then $A$ terminates the protocol.
Otherwise, from the information $A$ received via path $q$, $A$ decides
which path from $A$ to $B$ is corrupted and recovers $B$'s authentication
key $(a^A,b^A)$. $A$ sends $(M^A, \auth(M^A; a^A,b^A))$
to $B$ via the uncorrupted path from $A$ to $B$.
\item $B$ recovers the message and checks that the authenticator
is correct.
\end{algorithm}
Similarly as in the proof of Theorem \ref{fwsuf}, it can be shown that the
above protocol is $(0,\delta)$-secure against a $1$-active adversary.
\hfill{Q.E.D.}\end{proof}

\begin{theorem}
\label{probsufficient}
Let $G(V,E)$ be a directed graph, $A,B\in V$, and $k\ge u\ge 1$. If there are
$2k+1-u$ directed node disjoint paths $p_1$, $\ldots$, $p_{2k+1-u}$ 
from $A$ to $B$, and $u$ directed node disjoint 
paths $q_1$,   $\ldots$, $q_u$
($q_1$,   $\ldots$, $q_u$ are node disjoint from 
$p_1$, $\ldots$, $p_{2k+1-u}$) from $B$ to $A$, 
then for any $0<\delta<\frac{1}{2}$, there is an efficient
$(0,\delta)$-secure message transmission protocol from $A$ to $B$ 
against a $k$-active adversary.
\end{theorem}

Before we give an efficient $(0,\delta)$-secure 
message transmission protocol from $A$ to $B$. We first demonstrate the
underlying idea by giving a non-efficient  (exponential in $k$) 
$(0,\delta)$-secure 
message transmission protocol from $A$ to $B$ against a $k$-active adversary. 
Let $M^A\in {\bf F}$ be 
the secret that $A$ wants to send to $B$, and ${\cal P}_1, \ldots,
{\cal P}_t$ be an enumeration of size $k+1$ 
subsets of $\{p_1,\ldots, p_{2k+1-u}, q_1,\ldots, q_u\}$. The 
protocol proceeds from round $1$ through $t$. In each round $1\le m\le t$,
we have the following steps:
\begin{algorithm}{9}
\item For each $p_i\in {\cal P}_m$, $A$ chooses 
$(a^A_{i,m}, b^A_{i,m})\in_R{\bf F}^2$ and sends
$(a^A_{i,m}, b^A_{i,m})$ to $B$ via $p_i$.
\item For each $p_i\in {\cal P}_m$, $B$ receives 
$(a^B_{i,m}, b^B_{i,m})$ from $A$ via $p_i$.
\item For each $q_i\in {\cal P}_m$, $B$ chooses 
$(c^B_{i,m}, d^B_{i,m})\in_R{\bf F}^2$ and sends
$(c^B_{i,m}, d^B_{i,m})$ to $A$ via $q_i$.
\item For each $q_i\in {\cal P}_m$, $A$ receives 
$(c^A_{i,m}, d^A_{i,m})$ from $B$ via $q_i$.
\item $A$ computes $C^A=\sum_{p_i\in{\cal P}_m}a^A_{i,m}+
\sum_{q_i\in{\cal P}_m}c^A_{i,m}$,
$D^A=\sum_{p_i\in{\cal P}_m}b^A_{i,m}+
\sum_{q_i\in{\cal P}_m}d^A_{i,m}$,
and sends $(M^A+C^A, \auth(M^A+C^A; C^A,D^A))$ to $B$ via all paths
in $p_i$ in  ${\cal P}_m$.
\item For each $p_i\in {\cal P}_m$,
$B$ receives $(e^B_{i,m},f^B_{i,m})$ from $A$ via $p_i$.
\item If $(e^B_{i,m},f^B_{i,m})=(e^B_{j,m},f^B_{j,m})$ for all 
$p_i, p_j\in {\cal P}_m$, then $B$ goes to Step \ref{expnn1}. Otherwise,
$B$ goes to round $m+1$.
\item\label{expnn1} $B$ computes $C^B=\sum_{p_i\in{\cal P}_m}a^B_{i,m}+
\sum_{q_i\in{\cal P}_m}c^B_{i,m}$,
$D^B=\sum_{p_i\in{\cal P}_m}b^B_{i,m}+
\sum_{q_i\in{\cal P}_m}d^B_{i,m}$.
\item If $f^B_{i,m}=\auth(f^B_{i,m}; C^B,D^B)$, then $B$ computes the
secret $M^B=f^B_{i,m}-C^B$ and terminates the protocol. Otherwise,
$B$ goes to round $m+1$.
\end{algorithm}
Since there is at least one ${\cal P}_m$ such that all paths 
in ${\cal P}_m$ are not corrupted, $B$ receives the correct 
secret by the end of the protocol with high probability. 
It is also straightforward to check that
the above protocol has perfect secrecy.

\begin{proof} (Proof of Theorem \ref{probsufficient})
Let $M^A\in {\bf F}$ be the secret that $A$ wants to
send to $B$. $A$ constructs $(k+1)$-out-of-$(2k+1-u)$ secret shares
$(s_1^A, \ldots, s_{2k+1-u}^A)$ of $M^A$. The protocol proceeds
from round $1$ through $2k+2+u$. For each round $1\le i\le 2k+1-u$,
we have the following steps:
\begin{algorithm}{9}
\item $A$ chooses $\{(a_{i,j}^A, b_{i,j}^A)\in_R{\bf F}^2$, 
$: 1\le j\le 2k+1-u\}$.
\item $A$ sends $\{s_i^A,
\auth(s_i^A; a_{i,1}^A,b_{i,1}^A)$, 
$\ldots$,
$\auth(s_i^A; a_{i,2k+1-u}^A,b_{i,2k+1-u}^A)\}$
to $B$ via path $p_i$, and sends  $(a^A_{i,j}, b_{i,j}^A)$
to $B$ via path $p_j$ for each $1\le j\le 2k+1-u$.
\item $B$ receives $\{s^B_i, 
d^B_{i,1}, \ldots, d^B_{i,2k+1-u}\}$ via path $p_i$, and
$(a_{i,j}^B, b_{i,j}^B)$ via path $p_j$ for each
$1\le j\le 2k+1-u$.
\item $B$ computes $t=|\{j: d_{i,j}^B=\auth(s_i^B; 
a_{i,j}^B, b_{i,j}^B)\}|$.
If $t\geq k+1$, then $B$ decides that $s_i^B$ is a valid share.
Otherwise $B$ decides that $s_i^B$ is an invalid share. 
\end{algorithm}
At the end of round $2k+1-u$, if $B$ has received $k+1$ valid shares, 
then $B$ recovers the secret $M^B$ from these valid shares
and terminates the protocol. Otherwise, $B$ proceeds to
round $2k+2-u$.  In round $2k+2-u$, we have the following steps:
\begin{algorithm}{9}
\item $A$ chooses $\{(a_i^A, b_i^A,c_i^A)\in_R{\bf F}^3$
$: 1\le i\le 2k+1-u\}$, and sends $(a_i^A, b_i^A,c_i^A)$ to $B$ via path 
$p_i$ for each $i\le  2k+1-u$.
\item For each $1\le i\le 2k+1-u$, 
$B$ receives  $(a_i^B, b_i^B,c_i^B)$ on path $p_i$ from $A$ (if no value
is received on path $p_i$, $B$ sets it to a default value).
\item For each $1\le i\le 2k+1-u$, $B$ chooses $r_i^B\in_R{\bf F}$ and
computes $\beta^B=\{(r_i^B, \auth(r_i^B; a_i^B, b_i^B,c_i^B)): 
1\le i\le 2k+1-u\}$.
\end{algorithm}
In each round $2k+3-u\le i\le 2k+2$, we have the following steps:
\begin{algorithm}{9}
\item $B$ chooses $(d_i^B, e_i^B)\in_R{\bf F}^2$ and
$\{(v_{i,j}^B, w_{i,j}^B)\in_R{\bf F}^2: 1\le j\le u\}$.
\item $B$ sends $(d_i^B, e_i^B)$, $\beta^B$,
and $\{\auth(\langle d_i^B, e_i^B\rangle; v_{i,j}^B, w_{i,j}^B)$
$: 1\le j\le u\}$ to $A$ via path $q_i$, and $(v_{i,j}^B, w_{i,j}^B)$
to $A$ via path $q_j$ for each $1\le j\le u$.
\item $A$ receives (or substitutes default values) 
$(d_i^A, e_i^A)$, $\beta_i^A$,
and $\{\alpha^A_{i,j}: 1\le j\le u\}$ from $B$ via path $q_i$, 
and $(v_{i,j}^A, w_{i,j}^A)$
from $B$ via path $q_j$ for each $1\le j\le u$.
\end{algorithm}
According to the values that $A$ has received,
$A$ divides the paths set $\{q_1,\ldots, q_u\}$ 
into subsets ${\cal Q}_1, \ldots, {\cal Q}_t$ such that for any
$l,m,n$ with $1\le l\le t$, $1\le m,n\le u$, and $q_m,q_n\in {\cal Q}_l$,
we have 
\begin{enumerate}
\item $\beta^A_m=\beta^A_n$;
\item $\alpha^A_{m,n}= 
\auth(\langle d_m^B, e_m^B\rangle; v_{m,n}^B, w_{m,n}^B)$;
\item $\alpha^A_{n,m}= 
\auth(\langle d_n^B, e_n^B\rangle; v_{n,m}^B, w_{n,m}^B)$.
\end{enumerate}
For each ${\cal Q}_l$, let $q_m\in {\cal Q}_l$ and
$\beta^A_m=\{(r_{i,l}^A,\gamma^A_{i,l}): 1\le i\le 2k+1-u\}$.
$A$ computes the number 
$$t_l=|\{i: \gamma^A_{i,l}=\auth(r_{i,l}^A; a^A_i,b^A_i,c^A_i), 
1\le i\le 2k+1-u\}|+|{\cal Q}_l|$$
If $t_l\le k$, then $A$ decides that ${\cal Q}_l$
is an unacceptable set, otherwise, $A$ decides that ${\cal Q}_l$
is an acceptable set. Let ${\cal Q}_l=\emptyset$ for $t<l\le u$.

For each round $2k+3\le l\le 2k+2+u$, we have the following steps:
\begin{algorithm}{9}
\item If  ${\cal Q}_l=\emptyset$  or  ${\cal Q}_l$ is an unacceptable set,
then go to round $l+1$.
\item $A$ computes ${\cal P}_l=\{p_i:
\gamma^A_{i,l}=\auth(r_{i,l}^A; a^A_i,b^A_i,c^A_i), 1\le i\le 2k+1-u\}$,
$C_l^A=\sum_{p_i\in {\cal P}_l}a^A_i+\sum_{q_i\in{\cal Q}_l}d^A_i$, and
$D_l^A=\sum_{p_i\in {\cal P}_l}b^A_i+\sum_{q_i\in{\cal Q}_l}e^A_i$.
\item $A$ sends $(\langle {\cal Q}_l, {\cal P}_l, M^A+C_l^A\rangle, 
\auth(\langle {\cal Q}_l, {\cal P}_l, M^A+C_l^A\rangle;
C_l^A,D_l^A))$ to $B$ via all
paths $p_i\in{\cal P}_l$.
\item $B$ receives $(\alpha^B_{i,l}, \beta^B_{i,l})$ from path $p_i$ for 
$1\le i\le 2k+1-u$.
\item For each $1\le i\le 2k+1-u$, $B$ computes 
$\alpha^B_{i,l}=\langle {\cal Q}^B_{i,l}, {\cal P}^B_{i,l}, 
\beta^B_{i,l}\rangle$, 
$C_{i,l}^B=\sum_{p_j\in {\cal P}_{i,l}}a^B_j+
\sum_{q_j\in{\cal Q}_{i,l}}d^B_j$, and
$D_{i,l}^B=\sum_{p_j\in {\cal P}_{i,l}}b^B_j+
\sum_{q_j\in{\cal Q}_{i,l}}e^B_j$.
\item For each  $1\le i\le 2k+1-u$, $B$ checks whether 
$\beta^B_{i,l}=\auth(\alpha^B_{i,l}; C_{i,l}^B, D_{i,l}^B)$.
If the equation holds, then $B$ computes the secret 
$M^B=\beta^B_{i,l}-C_{i,l}^B$.
\end{algorithm}
If $B$ has not got the secret at the end of round $2k+1-u$, then 
there exists an uncorrupted path $q_j$ from $B$ to $A$ and
a paths set ${\cal Q}_l$ such that $q_j\in {\cal Q}_l$ and the information
that $A$ receives from paths in ${\cal Q}_l$ are reliable.
Thus, at the end of round $2k+2+u$, $B$ will output a secret $M^B$.
It is easy to check that, with high probability, 
this secret is the same as $M^A$. 

It is straightforward to show that the protocol achieves perfect 
privacy. Thus it is a $(0,\delta)$-secure message transmission
transmission protocol from $A$ to $B$ against a $k$-active adversary.
\hfill{Q.E.D.}\end{proof}

\section{$(0,0)$-Secure message transmission in directed graphs}
\label{digraphsec}
In the previous section, we addressed probabilistic reliable
message transmission in directed graphs. In this section, we consider
perfectly reliable message transmission in directed graphs.
We will show that if there are $u$ 
directed node disjoint paths from $B$ to $A$,
then a necessary and sufficient condition for $(0,0)$-secure 
message transmission from $A$ to $B$  against a $k$-active adversary 
is that there are $\max\{3k+1-2u, 2k+1\}$ 
directed node disjoint paths from $A$ to $B$.

\begin{theorem}
\label{00necessary}
Let $G(V,E)$ be a directed graph, $A,B\in V$.
Assume that there are $u$ directed node disjoint paths from
$B$ to $A$. Then a necessary condition for $(0,0)$-secure 
message transmission from $A$ to $B$  against a $k$-active adversary 
is that there are $\max\{3k+1-2u, 2k+1\}$ 
directed node disjoint paths from $A$ to $B$.
\end{theorem}
\begin{proof}
If $u=0$, then by the results in \cite{ddwy}, we need $3k+1$ directed node
disjoint paths from $A$ to $B$ for $(0,0)$-secure message transmission
against a $k$-active adversary. 
If $u\ge\lceil\frac{k}{2}\rceil$, then again by the results 
in \cite{ddwy}, we need $2k+1$ 
directed node disjoint paths from $A$ to $B$ for $0$-reliable (that is,
perfectly reliable) message transmission from $A$ to $B$ 
against a $k$-active adversary.
From now on, we assume that $0<u< \lceil\frac{k}{2}\rceil$. 

For a contradiction,
we assume that there are only $3k-2u$ 
directed node disjoint paths from $A$ to $B$, 
denoted as $p_1,\ldots, p_{3k-2u}$. Let $q_1,\ldots, q_u$ be the directed
node disjoint paths from $B$ to $A$.

Let $\Pi$ be a $(0,0)$-secure message transmission protocol 
from $A$ to $B$. In the following, we will construct a $k$-active 
adversary to defeat this protocol. The transcripts
distribution $view^A_\Pi$ of $A$ is drawn from a probability distribution that
depends on the message $M^A$ to be transmitted by $A$, 
the coin flips $R^A$ of $A$,
the coin flips $R^B$ of $B$, the coin flips $R^{\cal A}$ of the adversary 
(without loss of generality, we assume that the value $R^{\cal A}$ will 
determine the choice of faulty paths controlled by the adversary), 
and the coin flips $R^{\cal T}$ of
all other honest nodes. Without loss of generality, we can 
assume that the protocol
proceeds in steps, where $A$ is silent during even steps and $B$ is silent 
during odd steps (see \cite{ddwy}).

The strategy of the adversary is as follows. 
First it uses $R^{\cal A}$ to choose a value $b$. If $b=0$,
then it uses $R^{\cal A}$ again to choose $k$ directed paths 
$p_{a_1}, \ldots, p_{a_k}$ from $A$ to $B$ and controls the first node on each
of these $k$ paths. If $b=1$, then it uses $R^{\cal A}$ again to choose $k-u$ 
directed paths $p_{a_1}, \ldots, p_{a_{k-u}}$ from $A$ to $B$ 
and controls the first node on each of these $k-u$ paths
and the first node on each of the $u$ paths from $B$ to $A$. 
It also  uses $R^{\cal A}$ to choose a message $\hat{M}^A\in {\bf F}$
according to the same probability distribution from which the actual message
$M^A$ was drawn. In the following we describe the protocol the adversary 
will follow.
\begin{itemize}
\item Case $b=0$. The $k$ paths $p_{a_1}, \ldots, p_{a_k}$ 
behaves as a passive adversary. That is, it proceeds according to the
protocol $\Pi$. 
\item Case $b=1$. The $k-u$ paths $p_{a_1}, \ldots, p_{a_{k-u}}$ 
ignores what $A$ sends in
each step of the protocol and simulates what $A$ would send to $B$ when
$A$ sending $\hat{M}^A$ to $B$. The $u$ paths from $B$ ignores 
what $B$ sends in each step
of the protocol and simulates what $B$ would send to $A$ when $b=0$.
\end{itemize}
In the following, we assume that the tuple 
$(M^A, R^A,R^B, R^{\cal T}, R^{\cal A})$ is fixed, $b=0$,
the protocol halts in $l$ steps, and the view of $A$ is
$view_\Pi^A(M^A, R^A,R^B, R^{\cal T}, R^{\cal A})$. Let
$\alpha^A_{i,j}$ be the values that $A$ sends on path $p_i$
in step $j$ and $\vec{\alpha}^A_i =(\alpha^A_{i,1}, \ldots, \alpha^A_{i,l})$.
We can view $\vec{\alpha}^A_i$ as shares of the message $M^A$.
Similarly, let $\alpha^B_{i,j}$ be the values that $B$ receives on path $p_i$
in step $j$ and $\vec{\alpha}^B_i =(\alpha^B_{i,1}, \ldots, \alpha^B_{i,l})$.

First, it is straightforward to show that for any $k$ paths
$p_{a_1}, \ldots, p_{a_k}$ from $A$ to $B$,
there is an $\hat{R}^{\cal A}_1$ such that $b=0$,
the adversary controls the paths $p_{a_1}, \ldots, p_{a_k}$,  and 
\begin{equation}
\label{indis00}view_\Pi^A(M^A, R^A,R^B, R^{\cal T}, R^{\cal A})=
view_\Pi^A(M^A, R^A,R^B, R^{\cal T}, \hat{R}^{\cal A}_1)
\end{equation}
Due to the fact that $\Pi$ is a perfectly private message 
transmission protocol, from any $k$ shares from
$(\vec{\alpha}^A_1, \vec{\alpha}^A_2,\ldots, \vec{\alpha}^A_{3k-2u})$ one
cannot recover the secret message $M^A$. Thus 
$(\vec{\alpha}^A_1, \vec{\alpha}^A_2,\ldots, \vec{\alpha}^A_{3k-2u})$
is at least a $(k+1)$-out-of-$(3k-2u)$ secret sharing scheme.

Secondly, for any $k-u$ paths
$p_{a_1}, \ldots, p_{a_{k-u}}$ from $A$ to $B$,
there is an $\hat{R}^{\cal A}_2$ such that $b=1$, $\hat{M}^A\not=M^A$,
the adversary controls the paths 
$p_{a_1}, \ldots, p_{a_{k-u}}, q_1,\ldots, q_u$,  and 
\begin{equation}
\label{indis01}view_\Pi^A(M^A, R^A,R^B, R^{\cal T}, R^{\cal A})=
view_\Pi^A(M^A, R^A,R^B, R^{\cal T}, \hat{R}^{\cal A}_2)
\end{equation}
Due to the fact that $\Pi$ is a perfectly reliable message 
transmission protocol, any $k-u$ errors in the shares
$(\vec{\alpha}^B_1, \vec{\alpha}^B_2,\ldots, \vec{\alpha}^B_{3k-2u})$
can be corrected by $B$ to recover the secret message $M^A$.

In summary, 
$(\vec{\alpha}^A_1, \vec{\alpha}^A_2,\ldots, \vec{\alpha}^A_{3k-2u})$
is at least a $(k+1)$-out-of-$(3k-2u)$ secret sharing scheme
that can correct $k-u$ errors.
By the results in \cite{MacWilliamsSloane78}, we know that the maximum
number of errors that a $(k+1)$-out-of-$(3k-2u)$ secret sharing scheme
could correct is
$$\left\lfloor\frac{(3k-2u)-k-1}{2}\right\rfloor =
\left\lfloor\frac{2k-2u-1}{2}\right\rfloor=k-u-1.$$
This is a contradiction, which concludes the proof.
\hfill{Q.E.D.}\end{proof}

For the sufficient condition, we first show the simple case for $u=1$.
\begin{theorem}
\label{perfectu1}
Let $G(V,E)$ be a directed graph, $A,B\in V$, and $k\ge 2$.
If there are $3k-1$ directed node disjoint paths $p_1,\ldots, p_{3k-1}$ from
$A$ to $B$ and one directed path $q$ from $B$ to $A$ ($q$ is node 
disjoint from $p_1,\ldots, p_{3k-1}$) then there is a 
$(0,0)$-secure message transmission
protocol  from $A$ to $B$ against a $k$-active adversary.
\end{theorem}
\begin{proof} 
In the following protocol $\pi$, $A$ $(0,0)$-securely
transmits  $M^A\in_R{\bf F}$ to $B$: 
\begin{algorithm}{9}
\item Both $A$ and $B$ sets $I=1$.
\item\label{u11} $A$ constructs $(k+1)$-out-of-$(3k-1)$ MDS secret 
shares $(s^A_{1,I}, ..., s^A_{3k-1,I})$
of $M^A$. For each $1\le i\le 3k-1$, $A$ sends $s^A_{i,I}$ to $B$ via the
path $p_i$.
\item\label{bsends} For each $i\le 3k-1$, $B$ receives $s^B_{i,I}$
on path $p_i$. By correcting at most $k-1$ errors, $B$ recovers a value
$\hat{M}_I^B$ from these shares. $B$ sends $s^B_{I,I}$ to $A$ via the path $q$.
\item\label{areceives} $A$ receives $\bar{s}_{I,I}^A$ from $q$.
$A$ distinguishes the following two cases:
\begin{enumerate}
\item $\bar{s}^A_{I,I}=s^A_{I,I}$. $A$ reliably sends  ``path $p_I$ maybe OK'',
sets $I=I+1$. If $I>3k-1$ then goes to Step \ref{00final}, otherwise,
goes to Step \ref{u11}.
\item $\bar{s}^A_{I,I}\not=s^A_{I,I}$. $A$ reliably sends  
``path $p_I$ or $q$ is faulty''. $A$ constructs $k$-out-of-$(3k-2)$ MDS secret 
shares $\{r^A_{i,I}: i\not= I, 1\le i\le 3k-1\}$
of $M^A$. For each $i\le 3k-1$ such that $i\not=I$, 
$A$ sends $r^A_{i,I}$ to $B$ via the path $p_i$. $A$ terminates the protocol.
\end{enumerate}
\item\label{brecovers} $B$ distinguishes the following two cases:
\begin{enumerate}
\item $B$ reliably receives ``path $p_I$ maybe OK''. 
$B$ sets $I=I+1$. If $I>3k-1$ then goes to Step \ref{00final}, otherwise,
goes to Step \ref{u11}.
\item $B$ reliably receives ``path $p_I$ or $q$ is faulty''.
In this case, $B$ also receives  $r^B_{i,I}$ from path $p_i$ for each 
$i\not=I$. By correcting at most $k-1$ errors, 
$B$ recovers $M^B$ from these shares and terminates the protocol.
\end{enumerate}
\item\label{00final} $B$ checks whether $\hat{M}^B_i=\hat{M}^B_j$
for all $1\le i,j \le 3k-1$. If all these values are equal,
then $B$ sets $M^B=\hat{M}^B_1$, sends ``stop'' to $A$ via $q$,
and terminates the protocol. Otherwise, $B$ sends the shares 
$(s^B_{1,3k-1}, ..., s^B_{3k-1,3k-1})$ to $A$ via $q$.
\item $A$ distinguishes the following two cases:
\begin{enumerate}
\item $A$ receives 
$(\bar{s}^A_{1,3k-1}, ..., \bar{s}^A_{3k-1,3k-1})$ on the path $q$.
$A$ computes ${\cal P}=\{i: \bar{s}^A_{i,3k-1}\not= s^A_{i,3k-1}\}$, 
reliably sends ${\cal P}$ to $B$, and terminates the protocol.
\item $A$ receives anything else. $A$ terminates the protocol.
\end{enumerate}
\item $B$ reliable receives ${\cal P}$ from $A$, recovers 
$M^B$ from the shares $\{s^B_{i,3k-1}: i\notin{\cal P}\}$, and 
terminates the protocol.
\end{algorithm}
Since there could be $k$ faulty paths from $A$ to $B$, and a
$(k+1)$-out-of-$(3k-1)$ MDS secret sharing scheme can 
correct at most $k-1$ errors and simultaneously detect $k-1$ errors. 
$B$ may recover an incorrect message $\hat{M}^B_I$ in Step \ref{bsends}.
$B$ therefore needs to verify whether it has recovered the correct
message in the following steps. Note that if $q$ is faulty, then $B$ must
have recovered the correct message in Step  \ref{bsends} for each $I$.
In Step  \ref{bsends}, $B$ also sends $s^B_{I,I}$ to $A$ via
the path $q$. This does not violate the perfect privacy property since
if there are $k-1$ faulty paths from $A$ to $B$, then the adversary
gets at most $k$ shares including this share, and  if there are 
$k$ faulty paths from $A$ to $B$, then the adversary does not 
control the path $q$ and does not get this share.

In Step \ref{areceives}, if $\bar{s}^A_{I,I}=s^A_{I,I}$,
then it could be the case that $s^B_{I,I}=s^A_{I,I}$
or it could be the case that the path $q$ is faulty. In any case, 
we have to continue the protocol further. However, if 
$\bar{s}^A_{I,I}\not=s^A_{I,I}$, then $A$ is convinced that 
either $p_I$ or $q$ is faulty. Since a
$k$-out-of-$(3k-2)$ MDS secret sharing scheme could correct $k-1$ errors
and detect $k-1$ errors simultaneously. $B$ will recover the correct message
in Step \ref{brecovers}.

Now assume that $B$ does not recover the correct message at the end 
of round $I=3k-1$. We can distinguish the following two cases:
\begin{itemize}
\item $B$ recovers the same message $\hat{M}^B_I$ in all 
$3k-1$ rounds. If this happens, $B$ is convinced that
this uniquely recovered message is the correct message.
Note that this follows from the following two arguments: 
If the path $q$ is faulty,
then obviously $B$ has recovered the same correct message in each round.
If the path $q$ is non-faulty and we assume that the path 
$p_t$ ($1\le t\le 3k-1$) is faulty, then in order for the adversary 
to avoid being caught by $A$ in round $t$, $p_t$ must behave
nicely in round $t$, that is, $s^A_{t,t}=s^B_{t,t}$ 
(otherwise $A$ has identified that 
$q$ or $p_t$ is faulty in round $t$). Thus there are at most 
$k-1$ errors in the shares that $B$ received in round $t$ and 
$B$ recovers the correct message in round $t$.
\item $B$ recovers different messages in these  $3k-1$ rounds.
This happens only if $q$ is honest. Thus, $B$ could send
the shares it receives in round $3k-1$ to $A$ via path $q$ and 
$A$ can tell $B$ which shares are incorrect. Thus $B$ could recover the
correct message from these non-faulty shares (there are at least $2k-1$ 
non-faulty shares).
\end{itemize}
The above arguments show that the protocol $\pi$ is $(0,0)$-secure against
a $k$-active adversary.
\hfill{Q.E.D.}\end{proof}

\begin{theorem}
\label{perfectugen}
Let $G(V,E)$ be a directed graph, $A,B\in V$, and $k\ge 2$. 
If there are $n=\max\{3k+1-2u, 2k+1\}$ directed node disjoint paths 
$p_1,\ldots, p_n$ from
$A$ to $B$ and $u$ directed path $q_1,\ldots, q_u$ from 
$B$ to $A$ ($q_1,\ldots, q_u$ are node disjoint from 
$p_1,\ldots, p_n$) then there is a 
$(0,0)$-secure message transmission
protocol  from $A$ to $B$ against a $k$-active adversary.
\end{theorem}
\begin{proof} 
For $u=1$ or $k=2$, the result is proved in Theorem \ref{perfectu1}. 
We prove the theorem by induction. Assume that  $u>1$, $k>2$, and
the theorem holds for $u-1$ and $k-1$. In the following, 
we show that the Theorem holds for $u$ and $k$ by induction.

Let ${\cal H}=\{h_I: h_I=\langle p_{I_1}, \ldots, p_{I_u}\rangle\}$ 
be the set of all ordered $u$-subsets of 
$\{p_1,\ldots, p_n\}$. Then $|{\cal H}|=\frac{n!}{(n-u)!}$.
In the following protocol $\pi$, $A$ $(0,0)$-securely
transmits  $M^A\in_R{\bf F}$ to $B$: 
\begin{algorithm}{19}
\item\label{firststepu} Both $A$ and $B$ set $I=1$.
\item\label{ggu11} $A$ constructs $(k+1)$-out-of-$n$ MDS secret 
shares $(s^A_{1,I}, ..., s^A_{n,I})$
of $M^A$. For each $i\le n$, $A$ sends $s^A_{i,I}$ to $B$ via the
path $p_i$.
\item\label{ggbsends} For each $i\le n$, 
$B$ receives (or sets default) $s^B_{i,I}$ on path $p_i$. By correcting 
at most $k-u$ errors, $B$ recovers a value $\hat{M}_I^B$ from these shares.
For each $i\le u$, $B$ sends $s^B_{I_i,I}$ to $A$ via the path $q_i$.
Note that we assume $h_I=\langle p_{I_1}, \ldots, p_{I_u}\rangle$ here.
\item\label{ggareceives} For each $i\le u$, $A$ receives 
(or sets default) $\bar{s}_{I_i,I}^A$ 
from $q_i$.
$A$ distinguishes the following two cases:
\begin{enumerate}
\item $\bar{s}_{I_i,I}^A =s^A_{I_i,I}$ for all $i\le u$. 
$A$ reliably sends  ``paths in $h_I$ maybe OK'',
sets $I=I+1$. If $I>|{\cal H}|$ then $A$ goes to Step \ref{Iends}, 
otherwise, goes to Step \ref{ggu11}.
\item $\bar{s}_{I_{i_0},I}^A \not=s^A_{I_{i_0},I}$ for some $i_0\le u$. 
$A$ reliably sends  ``path $p_{I_{i_0}}$ or $q_{i_0}$ is faulty''. 
$A$ goes to the $(0,0)$-secure message transmission protocol
against a $(k-1)$-active adversary on the paths 
$\{p_i: i\not= I_{i_0}\}\cup\{q_i: i\not= i_0\}$
to transmit $M^A$ to $B$ (here we use induction).
\end{enumerate}
\item\label{ggbrecovers} $B$ distinguishes the following two cases:
\begin{enumerate}
\item $B$ reliably receives  ``paths in $h_I$ maybe OK''. 
$B$ sets $I=I+1$. If $I>|{\cal H}|$ then goes to 
Step \ref{Iends}, otherwise,
goes to Step \ref{ggu11}.
\item $B$ reliably receives ``path $p_{I_{i_0}}$ or $q_{i_0}$ is faulty''.
In this case, $B$ goes to the $(0,0)$-secure message transmission
protocol against a $(k-1)$-active adversary 
on the paths  $\{p_i: i\not= I_{i_0}\}\cup\{q_i: i\not= i_0\}$
and receives the message $M^B$. 
\end{enumerate}
\item\label{Iends}  
$B$ computes 
whether $\hat{M}^B_i=\hat{M}^B_j$
for all $i,j \le |{\cal H}|$. If all these values are equal,
then $B$ sets $M^B=\hat{M}^B_1$, sends ``stop'' to $A$ via all
paths $q_i$, and terminates the protocol. 
Otherwise, $B$ goes to Step \ref{00ggfinal}.
\item\label{allstop} If $A$ receives ``stop'' on all paths $q_1,\ldots, q_u$,
then $A$ terminates the protocol, otherwise, $A$ goes to Step \ref{00ggfinal}.
\item\label{00ggfinal} $A$ chooses $R^A_1\in_R{\bf F}$,  constructs 
$(k+1)$-out-of-$n$ MDS secret 
shares $(s^A_1, \ldots, s^A_n)$ of $R^A_1$, and sends $s^A_i$
to $B$ via path $p_i$ for each $i\le n$.
\item\label{bdetect1} For each $i\le n$, $B$ receives 
(or sets default) $s^B_i$ on path $p_i$.
$B$ distinguishes the following two cases:
\begin{enumerate}
\item There are errors in the shares $(s^B_1, \ldots, s^B_n)$.
In this case, for each $j\le u$, $B$ sends  $(s^B_1, \ldots, s^B_n)$
to $A$ via the path $q_j$. 
Note that a $(k+1)$-out-of-$n$ MDS secret 
sharing scheme could be used to detect $n-k-1\ge k$ errors.
\item There is  no error in the shares $(s^B_1, \ldots, s^B_n)$. $B$
recovers the value $R_1^B$ from these shares and for each $j\le u$, $B$
sends ``OK'' to $A$ via path $q_j$.
\end{enumerate}
\item\label{Jareceive} For each $j\le u$, $A$ receives (or sets default)
$(\bar{s}_{1,j}^A,\ldots, \bar{s}_{n,j}^A)$ from the path
$q_j$. $A$ distinguishes the following two cases:
\begin{enumerate}
\item $\bar{s}_{i_0,j_0}^A \not=s_{i_0}^A$  for some $i_0\le n$ and 
$j_0\le u$. 
$A$ reliably sends  ``path $p_{i_0}$ or $q_{j_0}$ is faulty'' to $B$. 
$A$ goes to the $(0,0)$-secure message transmission protocol against
a $(k-1)$-active adversary
on the paths $\{p_i: i\not= i_0\}\cup\{q_j: j\not= j_0\}$
to transmit $M^A$ to $B$ (here we use induction again).
\item All other cases. $A$ reliably transmits  ``continue the protocol''
to $B$ and goes to Step \ref{sendsR2}.
\end{enumerate}
\item\label{Jbrecovers} $B$ distinguishes the following two cases:
\begin{enumerate}
\item $B$ reliably receives  ``continue the protocol''. 
$B$ goes to Step \ref{sendsR2}.
\item $B$ reliably receives ``path $p_{i_0}$ or $q_{j_0}$ is faulty''.
In this case, $B$ goes to the $(0,0)$-secure message transmission
protocol against a $(k-1)$-active adversary 
on the paths  $\{p_i: i\not= i_0\}\cup\{q_j: j\not= j_0\}$
and receives the message $M^B$. 
\end{enumerate}
\item\label{sendsR2} $A$ computes $R^A_2=M^A-R^A_1$,  constructs 
$(k+1)$-out-of-$n$ MDS secret 
shares $(s^A_1, \ldots, s^A_n)$ of $R^A_2$, and sends $s^A_i$
to $B$ via path $p_i$ for each $i\le n$.
\item\label{bdetect2} For each $i\le n$, $B$ receives 
(or sets default) $s^B_i$ on path $p_i$.
$B$ distinguishes the following two cases:
\begin{enumerate}
\item There are errors in the shares $(s^B_1, \ldots, s^B_n)$.
In this case, for each $j\le u$, $B$ sends  $(s^B_1, \ldots, s^B_n)$
to $A$ via the path $q_j$. 
\item There is no error in the shares $(s^B_1, \ldots, s^B_n)$. $B$
recovers the value $R_2^B$ from these shares,
computes the secret $M^B=R^B_1+R^B_2$, and for each $j\le u$, $B$
sends ``OK'' to $A$ via path $q_j$. $B$ terminates the protocol.
\end{enumerate}
\item\label{Jareceive2} For each $j\le u$, $A$ receives (or sets default)
$(\bar{s}_{1,j}^A,\ldots, \bar{s}_{u,j}^A)$ from the path
$q_j$. $A$ distinguishes the following two cases:
\begin{enumerate}
\item $\bar{s}_{i_0,j_0}^A \not=s_{i_0}^A$  for some $i_0\le n$ and 
$j_0\le u$. 
$A$ reliably sends  ``path $p_{i_0}$ or $q_{j_0}$ is faulty'' to $B$. 
$A$ goes to the $(0,0)$-secure message transmission protocol against
a $(k-1)$-active adversary
on the paths $\{p_i: i\not= i_0\}\cup\{q_j: j\not= j_0\}$
to transmit $M^A$ to $B$.
\item All other cases. $A$ reliably transmits  ``the protocol is complete''
to $B$ and terminates the protocol.
\end{enumerate}
\item\label{Jbrecovers2} $B$ distinguishes the following two cases:
\begin{enumerate}
\item $B$ reliably receives  ``the protocol is complete''. 
$B$ terminates the protocol.
\item $B$ reliably receives ``path $p_{i_0}$ or $q_{j_0}$ is faulty''.
In this case, $B$ goes to the $(0,0)$-secure message transmission
protocol against a $(k-1)$-active adversary 
on the paths  $\{p_i: i\not= i_0\}\cup\{q_j: j\not= j_0\}$
and receives the message $M^B$. 
\end{enumerate}
\end{algorithm}
Since there could be $k$ faulty paths from $A$ to $B$, and a
$(k+1)$-out-of-$n$ MDS secret sharing scheme in Steps \ref{ggu11} 
and \ref{ggbsends} can  correct at most $k-u$ errors and 
simultaneously detect $k-u$ errors. 
$B$ may recover an incorrect message $\hat{M}^B_I$ in 
Step \ref{ggbsends}.
$B$ therefore needs to verify whether it has recovered the correct
message in the following steps. Note that if all the paths from $B$ to $A$ are 
faulty, then $B$ must have recovered the correct message 
in Step  \ref{ggbsends} for each $I\le |{\cal H}|$.
In Step  \ref{ggbsends}, $B$ also sends $s^B_{I_i,I}$ to $A$ via
the path $q_i$. This will not violate the perfect privacy property since
if there are $t$ faulty paths from $B$ to $A$, then the adversary
gets $k-t$ shares from the $A$ to $B$ paths and $t$ 
shares from the $B$ to $A$ paths. That is,
the adversary gets at most $k$ shares.

In Step \ref{ggareceives}, if $\bar{s}^A_{I_i,I}=s^A_{I_i,I}$ for all $i\le u$,
then for each $i\le u$, it could be the case that $s^B_{I_i,I}=s^A_{I_i,I}$
or it could be the case that the path $q_i$ is faulty. In any case, 
we have to continue the protocol further. However, if 
$\bar{s}^A_{I_{i_0},I}\not=s^A_{I_{i_0},I}$ for some $i_0\le u$,
then $A$ is convinced that either $p_{I_{i_0}}$ or $q_{i_0}$ is faulty. 
Thus if we delete the two paths $p_{I_{i_0}}$ and $q_{i_0}$, we 
have at most $k-1$ unknown faulty paths, $n-1$ paths from $A$ to $B$,
and $u-1$ paths from $B$ to $A$. Since 
$$\begin{array}{lll}
\max\{3(k-1)+1 - 2(u-1), 2(k-1)+1\}& = & \max\{3k-2u-4, 2k-1\}\\
& \le & \max\{3k-2u, 2k\}\\
& = &n-1,\end{array}$$
there is (by induction) a $(0,0)$-secure message transmission 
protocol from $A$ to $B$ against a $(k-1)$-active adversary on the paths  
$\{p_i: i\not= I_{i_0}\}\cup \{q_i: i\not= i_0\}$, $B$ recovers the correct
message $M^B$ in Step \ref{ggbrecovers}.

Assume that $B$ does not recover the correct message at the beginning
of Step \ref{Iends}. If $B$ recovers the same value $\hat{M}^B_I$ 
in all the $|{\cal H}|$ rounds between Step \ref{ggu11} and 
Step \ref{ggbrecovers}, then $B$ is convinced that
this uniquely recovered value is the correct message. 
Note that this follows from the following arguments: 
\begin{itemize}
\item All paths $q_1,\ldots, q_u$ from $B$ to $A$ are faulty. In this case
$B$ obviously has recovered the correct message in each round.
\item There is non-faulty path from $B$ to $A$. In this case,
let $t\ge 1$, $q_{i_1}, \ldots, q_{i_t}$ be a list of all non-faulty paths 
from $B$ to $A$, $p_{j_1}, \ldots, p_{j_t}$ be faulty,
and $h_I=\langle p_{I_1}, \ldots, p_{I_u}\rangle$ with 
$I_{i_1}=j_1, \ldots, I_{i_t}=j_t$. If 
$(s^B_{1,I}, ..., s^B_{n,I})$ is the shares that $B$ receives 
in Step \ref{ggbsends} of round $I$, then $s^B_{{j_1},I}=s^A_{{j_1},I}$,
$\ldots$, $s^B_{{j_t},I}=s^A_{{j_t},I}$ (otherwise $A$ identifies that 
some $q_i$ or $p_j$ is faulty in round $I$). That is, there are at most
$k-u$ errors in the shares $(s^B_{1,I}, ..., s^B_{n,I})$, and $B$ 
recovers the correct secret message in round $I$.
\end{itemize}

Now assume that $B$ does not recover the correct message at the beginning
of Step \ref{Iends} and $B$ recovers different values in 
these $|{\cal H}|$ rounds between Step \ref{ggu11} and 
Step \ref{ggbrecovers}. If this happens, then there must be non-faulty
paths from $B$ to $A$.  In this case, both $A$ and $B$ 
continues the protocol from Step \ref{00ggfinal}.
During Step \ref{00ggfinal} and Step \ref{Jbrecovers2},
$A$ tries to send $R^A_1$ and $R^A_2$ to $B$ using the
$(k+1)$-out-of-$n$ MDS secret sharing scheme. Since  there are 
at most $k$ faulty paths,
and a $(k+1)$-out-of-$n$ MDS secret sharing scheme could be used to 
correct $0$ error and simultaneously detect at least
$k$ errors, any error in these shares could be detected by $B$
in Steps \ref{bdetect1} and \ref{bdetect2}. Since there are non-faulty
paths from $B$ to $A$, any errors in these shares will be reported
back $A$ via the non-faulty $B$ to $A$ paths. Thus $A$ initiates
a $(0,0)$-secure message transmission protocol against
a $(k-1)$-active adversary
on the paths $\{p_i: i\not= i_0\}\cup\{q_j: j\not= j_0\}$ in
Step \ref{Jareceive} or Step \ref{Jareceive2} and $B$ will receive the
secret. If any error occurs in these cases, $B$ reports either the
shares of $R^A_1$ or the shares of $R^A_2$ (but not both) 
to $A$ via the $B$ to $A$ paths. Thus we have 
achieved the perfect privacy here.
On the other hand, if there is no error in these shares
of $R^A_1$ and $R^A_2$, then $B$ recovers the correct secret 
$M^B=R^B_1+R^B_2$.

We therefore proved that the protocol $\pi$ is $(0,0)$-secure 
against a $k$-active adversary.
\hfill{Q.E.D.}\end{proof}

In Theorem \ref{perfectugen}, we have the restriction that $k\ge 2$.
In the following we show a sufficient condition which is applicable to
$k=1$.

\begin{theorem}
\label{eperfectu1}
Let $G(V,E)$ be a directed graph, $A,B\in V$.
If there are $3k$ directed node disjoint paths $p_1,\ldots, p_{3k}$ from
$A$ to $B$ and one directed path $q$ from $B$ to $A$ ($q$ is node 
disjoint from $p_1,\ldots, p_{3k}$) then there 
is a $(0,0)$-secure message transmission
protocol  from $A$ to $B$ against a $k$-active adversary.
\end{theorem}
\begin{proof}
In the following protocol $\pi$, $A$ $(0,0)$-securely transmits
$M^A\in_R{\bf F}$ to $B$.
\begin{algorithm}{9}
\item\label{eu11} $A$ constructs $(k+1)$-out-of-$3k$ MDS 
secret shares $v^A=(s^A_1, ..., s^A_{3k})$
of $M^A$. For each $1\le i\le 3k$, $A$ sends $s_i$ to $B$ via the
path $p_i$.
\item\label{ebsends} Let $v^B=(s^B_1, ..., s^B_{3k})$
be the shares $B$ receives.
If $B$ finds that there are at most $k-1$ errors, $B$ recovers
$M^B$ from the shares, sends ``stop'' to $A$ via the path $q$,
and terminates the protocol.
Otherwise there are $k$ errors. In this case $B$ sends $v^B$ back
to $A$ via the path $q$ (note that $q$ is an honest path in this case).
\item $A$ distinguishes the following two cases:
\begin{enumerate}
\item $A$ receives $\bar{v}^A=(\bar{s}^A_1, ..., \bar{s}^A_{3k})$ 
from the path $q$.
$A$ reliably sends ${\cal P}=\{i: s^A_i\not= \bar{s}^A_i\}$ to $B$.
\item $A$ received ``stop'' or anything else via $q$.
$A$ terminates the protocol.
\end{enumerate}
\item $B$ reliably receives ${\cal P}$ from $A$. 
$B$ recovers $M^B$ from the shares $\{s^B_i: i\notin {\cal P}\}$ and
terminates the protocol
(note that $|\{s^B_i: i\notin {\cal P}\}|=2k$).
\end{algorithm}
Note that if $B$ sends $v^B$ to $A$ in Step \ref{ebsends} then
$k$ paths from $A$ to $B$ are corrupted and the path $q$ is honest.
Thus the adversary will not learn $v^B$. If the
adversary controls the path $q$, then it may change the message
``stop'' to something else. In this case,
$A$ will not be able to identify the corrupted paths from $A$ to $B$.
However, since $B$ has already recovered the key, $B$ will just ignore
the next received message. It is straightforward to show that
the protocol is $(0,0)$-secure.
\hfill{Q.E.D.}\end{proof}

\section{Efficient $(0,0)$-secure message transmission in directed graphs}
In the previous section, we proved a necessary and sufficient condition
for $(0,0)$-secure message transmission from $A$ to $B$. Our protocols in
these proofs are not efficient (exponential in $k$). In this section,
we show that if there are totally $3k+1$ paths between $A$ and $B$, 
then there are efficient (linear in $u$) $(0,0)$-secure message 
transmission protocols from $A$ to $B$.

\begin{theorem}
\label{perfectu}
Let $G(V,E)$ be a directed graph, $A,B\in V$ and $k\ge u$.
If there are $n=3k+1-u$ directed node
disjoint paths $p_1,\ldots, p_n$ from
$A$ to $B$ and $u$ directed node disjoint paths 
$q_1,\ldots, q_u$ from $B$ to $A$ 
($q_1,\ldots, q_u$ are node disjoint from  $p_1,\ldots, p_n$)
then there is an efficient $(0,0)$-secure message transmission
protocol from $A$ to $B$  against a $k$-active adversary.
\end{theorem}
\begin{proof} 
If we replace the steps between Step \ref{firststepu} and Step \ref{allstop}
of the protocol $\pi$ in the proof of Theorem \ref{perfectugen}
with the following steps:
\begin{algorithm}{9}
\item\label{eff1}  $A$ constructs $(k+1)$-out-of-$n$ MDS secret 
shares $(s^A_1, ..., s^A_n)$ of $M^A$. 
For each $i\le n$, $A$ sends $s^A_i$ to $B$ via the
path $p_i$.
\item\label{eff2}  For each $i\le n$, 
$B$ receives (or sets default) $s^B_i$ on path $p_i$. 
If there are at most $k-u$ errors in the shares
$(s^B_1, ..., s^B_n)$, then $B$ recovers the secret 
message $M^B$ from these shares by correcting the errors,
sends ``stop'' to $A$ via all paths $q_i$, and terminates the protocol. 
Otherwise, $B$ sends ``continue the protocol'' to $A$ via all paths $q_i$
and goes to Step \ref{00ggfinal} of the protocol $\pi$.
\item\label{eff3} If $A$ receives ``stop'' on all paths $q_1,\ldots, q_u$,
then $A$ terminates the protocol, otherwise, $A$ goes to Step \ref{00ggfinal}
of the protocol $\pi$.
\end{algorithm}
Note that a $(k+1)$-out-of-$n$ MDS secret 
sharing scheme could be used to detect $k$ errors and simultaneously
correct $k-u$ errors. Thus if all the paths $q_1,\ldots, q_u$ are 
controlled by the adversary, then $B$ recovers the secret message 
$M^B$ in Step \ref{eff2}. If at least one path from $B$ to $A$ is
not controlled by the adversary, then the protocol $\pi$ 
in the proof of Theorem \ref{perfectugen} starting from
Step \ref{00ggfinal} will let $B$ to recover the secret
message $M^B$. Here we should also note that the induction 
initiated in Step \ref{Jareceive} or Step \ref{Jareceive2} of 
the protocol $\pi$ works since $3k+1-u-1=3k-u> 3(k-1)+1-(u-1)$. 
It is straightforward that the protocol will
terminates in at most $11u$ steps.
\hfill{Q.E.D.}\end{proof}
In the previous theorems, including Theorem \ref{perfectu},
we have the restriction that the directed paths from $B$ to $A$ are all
node disjoint from the directed paths from $A$ to $B$.
In the following theorem we partially remove this restriction.
\begin{theorem}
\label{perfectustr}
Let $G(V,E)$ be a directed graph, $A,B\in V$.
Assume that there are $n=3k+1-u\ge 2k+1$ (which implies $k\ge u$)
directed node disjoint paths $p_1,\ldots, p_n$ from $A$ to $B$ 
and $u$ node disjoint directed paths $q_1,\ldots, q_u$ from $B$ to $A$. 
If $3k+1-2u$ paths among these $3k+1-u$
paths from $A$ to $B$ are node disjoint from the $u$ paths from $B$ to $A$,
then there is an efficient $(0,0)$-secure message transmission
protocol from $A$ to $B$  against a $k$-active adversary.
\end{theorem}

\begin{proof} 
We note that the proof of Theorem \ref{perfectu} 
could not be used here since if we remove (in the induction step) two 
paths $p_i$ and $q_j$ such that one of them is corrupted, 
we are not guaranteed that the $k$-active adversary becomes
a $(k-1)$-active adversary ($q_j$ may share a node with some other 
directed paths from $A$ to $B$ and that node could be corrupted).

First we describe the proof informally.
The protocol is divided into two phases.
In phase one of the protocol, $A$ tries to transmit
the secret message to $B$ assuming at least one of the directed paths
from $B$ to $A$ is not corrupted. This is done by running $u$ concurrent
sub-protocols in phase one, in each sub-protocol $B$ uses one of the 
directed paths from $B$ to $A$ to send some feedback information to $A$.
In the second phase of the protocol, $A$ transmits 
shares of the secret message through the $A$ to $B$ paths  excluding these
paths which have intersection with $B$ to $A$ paths. $B$ will use the 
information received in the second phase only if $B$ detects that all 
directed paths from $B$ to $A$ are corrupted in phase one.

In phase one, $A$ and $B$ execute the following protocol
on the path set $\{p_i:1\le i \le n\}\cup\{q\}$ for each directed path $q$ 
from $B$ to $A$. First $A$ chooses $R_0\in_R{\bf F}$ and
sends shares of $R_0$ to $B$ via the paths $p_1,\ldots, p_n$
using a $(k+1)$-out-of-$n$ MDS secret sharing scheme.
If $B$ can correct the errors in the received shares (that is, 
there were at most $k-u$ errors), $B$ recovers $R_0$.
Otherwise $B$ needs help from $A$ and $B$ sends the
received shares back to $A$ via the $B$ to $A$ path $q$.
The problems are that: $B$ may receive help even if $B$ has never asked for.
However $B$ can detect this. Therefore $B$ always works with $A$ on
such a protocol and recovers the correct $R_0$. 
Then $A$ sends $R_1=M^A-R_0$ using a $(k+1)$-out-of-$n$
MDS secret sharing scheme. If $B$ can correct the errors in the 
received shares  of $R_1$, $B$ has found the secret and can terminate the
protocol. If $B$ cannot correct these errors, $B$ needs to continue the
protocol. In this situation, $B$ distinguishes the following two cases:
\begin{enumerate}
\item $B$ has not asked for help in the transmission of $R_0$.
$B$ can ask for help now and $B$ will then recover the secret $M^A$.
\item $B$ has asked for help in the transmission of $R_0$. In this case $B$
cannot ask for help (otherwise the adversary may learn both
the values of $R_0$ and $R_1$ and thus may recover the secret).
The sub-protocol needs to be restarted (that is, $A$ constructs 
different $R_0$ and $R_1$ for $M^A$ and sends them to $B$ again). 
Each time when $A$ and $B$ restart this sub-protocol, $A$ sends 
the shares of $R_0$ and $R_1$ only via these ``non-corrupted'' paths 
from $A$ to $B$.
The ``non-corrupted'' paths are computed from the feedbacks that 
$A$ has received from the path $q$. If $q$ is not corrupted, then
the computation is reliable. However, if $q$ is corrupted, then
the computation is unreliable. If there is at least one
non-corrupted path $q_{i_0}$ from $B$ to $A$, then $B$ recovers the secret
from the sub-protocol running on the path set 
$\{p_1,\ldots, p_n\}\cup\{q_{i_0}\}$. Otherwise 
$B$ cannot recover the secret in phase one and we will go to phase two.
\end{enumerate}
If $B$ asks for help in the transmission of $R_0$,
then both $A$ and $B$ ``identify'' the corrupted paths from $A$ and $B$
according to the information that $B$ sends to $A$ via the path $q$.
If $k'$ dishonest paths from $A$ to $B$ have been (correctly or incorrectly) 
identified at the restart of the sub-protocol, $A$ uses 
a $(k+1)$-out-of-$(3k+1-u-k')$ MDS secret sharing scheme.
This MDS secret sharing scheme will only be used for error detection (or
message recovery in the case that no error occurs),
thus it can be used to detect $3k+1-u-k'-k -1= 2k-u-k'\ge k-k'$ errors.
Due to the fact that this MDS secret sharing scheme cannot
detect $k$ errors we need to organize ourselves that $B$ will
never use incorrectly identified paths from $A$ to $B$  since otherwise
$B$ could compute the incorrect ``secret". This is easy to be addressed
by having $B$ detect whether the path $q$ from $B$ to $A$ is dishonest or not.
This is done by having $A$ reliably send to $B$ what
$A$ received via the path $q$.
Since a $(k+1)$-out-of-$(3k+1-u)$ MDS secret sharing scheme
can detect $k$ errors and simultaneously correct $k-u$ errors,
both $A$ and $B$ identify at least $k'\geq k-u+1$
dishonest paths from $A$ to $B$ in the first run of the sub-protocol. 
During each following run of the sub-protocol, $B$ will either recover the
secret message (when no error occurs) or detect at least one corrupted path
from $A$ to $B$ ($A$ could also detect the corrupted path from $A$ to $B$
according to the information $A$ received on the path $q$). 
Thus the sub-protocol will be restarted at most $u$ times.

In phase two of the protocol, 
$A$ constructs $(k+1)$-out-of-$(3k+1-2u)$
MDS shares $(s_1,\ldots, s_{3k+1-2u})$ of the secret $M^A$
and sends these shares to $B$ via the  $3k+1-2u$ paths which are
node disjoint from the paths from the $u$ paths from $B$ to $A$.
Note that if $B$ has determined that all these $u$ paths from $B$ to $A$
have been corrupted in phase one, then $B$ recovers the secret
$M^A$ from the received shares $(s^B_1,\ldots, s^B_{3k+1-2u})$ in phase two
since a $(k+1)$-out-of-$(3k+1-2u)$ MDS secret sharing scheme
can be used to detect and correct $k-u$ errors simultaneously.
Note that if at least one path from $B$ to $A$ is honest in phase one, then
$B$ has recovered the secret in phase and can just ignore this last
message.

Now we present the entire protocol formally. 
\begin{algorithm}{99}
\item $B$ sets $\mathrm{BA\_BAD}=\emptyset$.
For each directed path $q$ from $B$ to $A$, $A$ and $B$ 
run the sub-protocol between Step \ref{begss} and 
Step \ref{phase1end} (the sup-protocols for the $u$ paths could be run 
parallely).
\item\label{begss} $A$ sets $\mathrm{AB\_CHANNEL}^A=\{p_1,\ldots, p_n\}$ 
and $j^A=0$. $B$ sets $\mathrm{AB\_CHANNEL}^B=\{p_1,\ldots, p_n\}$ 
and $j^B=0$. 
\item\label{begssj} Let $n_j=|\mathrm{AB\_CHANNEL}^A|$. 
$A$ chooses $R_0\in_R{\bf F}$, and 
constructs $(k+1)$-out-of-$n_j$ MDS secret
shares $\{s^A_i: p_i\in \mathrm{AB\_CHANNEL}^A\}$ of $R_0$.
For each $p_i\in \mathrm{AB\_CHANNEL}^A$, $A$ sends $s^A_i$ to
$B$ via the path $p_i$.
\item\label{bfim} For each $p_i\in \mathrm{AB\_CHANNEL}^B$, 
$B$ receives $s^B_i$ from 
$A$ via the path $p_i$. $B$ distinguishes the following two cases:
\begin{enumerate}
\item $B$ can recover $R_0$. If $j=0$ and there are 
at most $k-u$ errors, $B$ recovers $R_0$ by correcting the errors
(note that a $(k+1)$-out-of-$n$ MDS scheme can be used to detect $k$ errors
and simultaneously correct $k-u$ errors). 
If $j>0$, then $B$ recovers $R_0$ only if there is no error in 
the received shares. $B$ sends ``ok'' to $A$ via the path $q$.
\item $B$ cannot recover $R_0$. $B$ sends 
$\{s^B_i: p_i\in \mathrm{AB\_CHANNEL}^B\}$ to $A$ via the path $q$.
\end{enumerate}
\item $A$ distinguishes the following two cases:
\begin{enumerate}
\item $A$ receives ``ok'' via the path $q$. $A$ reliably sends
``ok'' to $B$.
\item $A$ receives $\{\bar{s}^B_i: p_i\in \mathrm{AB\_CHANNEL}^A\}$
(or sets default values if the received values are not in valid format).
$A$ sets $\mathrm{BAD}^A=\{p_i: \bar{s}^B_i\not= s^A_i\}$
and reliably sends $\{\bar{s}^B_i: p_i\in \mathrm{AB\_CHANNEL}^A\}$ 
and $\mathrm{BAD}^A$ to $B$. $A$ sets 
$\mathrm{AB\_CHANNEL}^A=\mathrm{AB\_CHANNEL}^A\setminus \mathrm{BAD}^A$, 
\end{enumerate}
\item $B$ distinguishes the following two cases:
\begin{enumerate}
\item $B$ reliably receives ``ok'' from $A$. If $B$ sent
``ok'' to $A$ in the Step \ref{bfim}, then goes to Step \ref{sendingr1}.
Otherwise, $B$ sets $\mathrm{BA\_BAD}=\mathrm{BA\_BAD}\cup\{q\}$
and goes to Step \ref{phase1end}.
\item $B$ reliably receives
$\{\bar{s}^B_i: p_i\in \mathrm{AB\_CHANNEL}^B\}$  and 
$\mathrm{BAD}^B$ from $A$.
If $\bar{s}^B_i=s^B_i$ for all  $p_i\in \mathrm{AB\_CHANNEL}^B$,
then $B$ sets $\mathrm{AB\_CHANNEL}^B=\mathrm{AB\_CHANNEL}^B\setminus 
\mathrm{BAD}^B$, recovers $R_0$ from 
$\{s^B_i: p_i\in \mathrm{AB\_CHANNEL}^B\}$, and goes to Step  \ref{sendingr1}.
Otherwise, $B$ sets $\mathrm{BA\_BAD}=\mathrm{BA\_BAD}\cup\{q\}$
and goes to Step \ref{phase1end}.
\end{enumerate}
\item\label{sendingr1} Let $n_j=|\mathrm{AB\_CHANNEL}^A|$. 
$A$ constructs $(k+1)$-out-of-$n_j$ MDS secret
shares $\{s^A_i: p_i\in \mathrm{AB\_CHANNEL}^A\}$ of $R_1=M^A-R_0$.
For each $p_i\in \mathrm{AB\_CHANNEL}^A$, $A$ sends $s^A_i$ to
$B$ via the path $p_i$.
\item\label{r1bfim} For each $p_i\in \mathrm{AB\_CHANNEL}^B$, 
$B$ receives $s^B_i$ from 
$A$ via the path $p_i$. $B$ distinguishes the following two cases:
\begin{enumerate}
\item $B$ can recover $R_1$. $B$ recovers $R_1$ only if there is no error in 
the received shares. $B$ sends ``ok'' to $A$ via the path $q$.
\item $B$ cannot recover $R_1$. For this situation we need
to distinguish two cases:
\begin{enumerate}
\item $B$ sent ``ok''  to $A$ in Step \ref{bfim}. That is,
$B$ has not asked for help to recover $R_0$. Then $B$ can ask for help now.
$B$ sends $\{s^B_i: p_i\in \mathrm{AB\_CHANNEL}^B\}$ to $A$ via the path $q$.
\item $B$ sent the received shares to $A$  in Step \ref{bfim}. That is,
$B$ has asked for help to recover $R_0$. Then $B$ cannot ask for help now.
$B$ sends ``continue to the next round'' to $A$ via the path $q$.
\end{enumerate}
\end{enumerate}
\item $A$ distinguishes the following three cases:
\begin{enumerate}
\item $A$ receives ``ok'' via the path $q$. $A$ reliably sends
``ok'' to $B$.
\item $A$ receives ``continue to the next round'' via the path $q$. 
$A$ sets $j^A=j^A+1$, reliably sends ``continue to the next round'' to $B$,
and goes to Step \ref{begssj}.
\item $A$ receives $\{\bar{s}^B_i: p_i\in \mathrm{AB\_CHANNEL}^A\}$
(or sets default values if the received values are in invalid format).
$A$ sets $\mathrm{BAD}^A=\{p_i: \bar{s}^B_i\not= s^A_i\}$,
$\mathrm{AB\_CHANNEL}^A=\mathrm{AB\_CHANNEL}^A\setminus \mathrm{BAD}^A$,
and reliably sends $\{\bar{s}^B_i: p_i\in \mathrm{AB\_CHANNEL}^A\}$ 
and $\mathrm{BAD}^A$ to $B$.
\end{enumerate}
\item $B$ distinguishes the following three cases:
\begin{enumerate}
\item $B$ reliably receives ``ok'' from $A$. If $B$ sent
``ok'' to $A$ in the Step \ref{r1bfim}, then 
$B$ has recovered the secret. $B$ terminates the entire protocol.
Otherwise, $B$ sets $\mathrm{BA\_BAD}=\mathrm{BA\_BAD}\cup\{q\}$
and goes to Step \ref{phase1end}.
\item $B$ reliably receives ``continues to the next round''.
If $B$ sent ``continues to the next round'' to $A$ in the 
Step \ref{r1bfim}, then $B$ sets $j^B=j^B+1$ and goes to Step \ref{begssj}.
Otherwise, $B$ sets $\mathrm{BA\_BAD}=\mathrm{BA\_BAD}\cup\{q\}$
and goes to Step \ref{phase1end}.
\item $B$  reliably receives 
$\{\bar{s}^B_i: p_i\in \mathrm{AB\_CHANNEL}^B\}$  and 
$\mathrm{BAD}^B$ from $A$.
If $\bar{s}^B_i=s^B_i$ for all  $p_i\in \mathrm{AB\_CHANNEL}^B$,
then $B$ sets $\mathrm{AB\_CHANNEL}^B=\mathrm{AB\_CHANNEL}^B\setminus 
\mathrm{BAD}^B$, recovers $R_1$ from 
$\{s^B_i: p_i\in \mathrm{AB\_CHANNEL}^B\}$, recovers
the secret $M^B$ from both $R_0$ and $R_1$, and terminates the entire protocol.
Otherwise, $B$ sets $\mathrm{BA\_BAD}=\mathrm{BA\_BAD}\cup\{q\}$
and goes to Step \ref{phase1end}.
\end{enumerate}
\item\label{phase1end} $B$ waits until all $u$ sub-protocols in phase one
finish. If $|\mathrm{BA\_BAD}|=u$ then $B$ goes to Step 
\ref{phase2}. Otherwise, $B$ has recovered the secret message, thus 
terminates the entire protocol.
\item\label{phase2} $A$ constructs $(k+1)$-out-of-$(3k+1-2u)$
MDS shares $(s_1,\ldots, s_{3k+1-2u})$ of the secret $M^A$
and sends these shares to $B$ via the  $3k+1-2u$ paths which are
node disjoint from the $u$ $B$ to $A$ paths.
Note that if $|\mathrm{BA\_BAD}|=u$, then $B$ can recover the 
secret message $M^B$ from the received shares 
$(s^B_1,\ldots, s^B_{3k+1-2u})$ since a $(k+1)$-out-of-$(3k+1-2u)$ MDS 
secret sharing scheme can be used to detect and correct $k-u$ errors 
simultaneously.
\end{algorithm}
It is straightforward to show that at the beginning of each run
of the sub-protocol between Step \ref{begss} and 
Step \ref{phase1end}, Both $A$ and $B$ have the same sets
of $\mathrm{AB\_CHANNEL}$, that is, $\mathrm{AB\_CHANNEL}^A
=\mathrm{AB\_CHANNEL}^B$ at Step \ref{begss}. From the analysis
before the above protocol, it is straightforward that the above protocol
is a $(0,0)$-secure message transmission protocol against a $k$-active
adversary.
\hfill{Q.E.D.}\end{proof}

\section{Secure message transmissions in hypergraphs}
\label{hypersec}
Applications of hypergraphs in secure communications have been
studied by Franklin and Yung in \cite{fy}.
A hypergraph $H$ is a pair $(V, E)$ where
$V$ is the node set and $E$ is the hyperedge set. Each
hyperedge $e\in E$ is a pair $(A, A^*)$ where $A\in V$ and
$A^*$ is a subset of $V$. In a  hypergraph, we assume that
any message sent by a node $A$ will be received
identically by all nodes in $A^*$, whether or not
$A$ is faulty, and all parties outside
of $A^*$ learn nothing about the content of
the message.

Let $A,B\in V$ be two nodes of the hypergraph $H(V,E)$.
We say that there is a ``{\em direct link}'' from node $A$ to
node $B$ if there exists a hyperedge $(A, A^*)$ such that
$B\in A^*$. We say that there is an ``{\em undirected link}''
from $A$ to $B$ if there is a directed link from $A$ to $B$
or a directed link from $B$ to $A$. If there is a directed (undirected)
link from $A_i$ to $A_{i+1}$ for every $i$, $0\le i<k$, then
we say that there is a ``{\em directed path}''
(``{\em undirected path}'') from
$A_0$ to $A_k$. $A$ and $B$ are ``{\em strongly $k$-connected}''
(``{\em weakly $k$-connected}'') in the hypergraph $H(V,E)$ if for all
$S\subset V-\{A,B\}$, $|S|<k$, there remains a directed (undirected)
path from $A$ to $B$ after the removal of $S$ and all hyperedges $(X,X^*)$
such that $S\cap (X^*\cup\{X\})\not=\emptyset$.
Franklin and Yung \cite{fy} showed that reliable and private
communication from $A$ to $B$ is possible against a $k$-{\it passive\/}
adversary if and only if $A$ and $B$ are strongly 1-connected
and weakly $(k+1)$-connected.
It should be noted that $B$ and $A$ are strongly $k$-connected
does not necessarily mean that $A$ and $B$ are strongly $k$-connected.

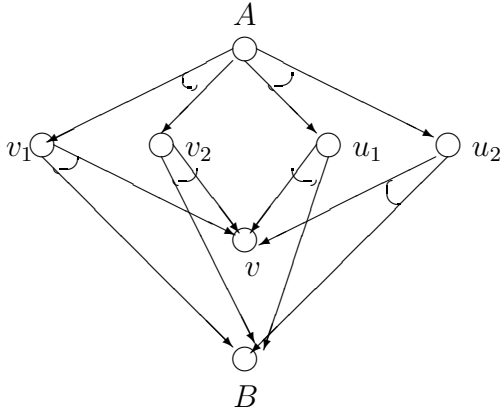
\begin{figure}[htb]
\begin{center}
\setlength{\unitlength}{3947sp}%
\begingroup\makeatletter\ifx\SetFigFont\undefined%
\gdef\SetFigFont#1#2#3#4#5{%
  \reset@font\fontsize{#1}{#2pt}%
  \fontfamily{#3}\fontseries{#4}\fontshape{#5}%
  \selectfont}%
\fi\endgroup%
\begin{picture}(2925,2535)(751,-1786)
\thinlines
{\put(1089,-286){\oval( 76, 12)[bl]}
\put(1089,-211){\oval(224,162)[br]}
}%
{\put(1876,277){\oval( 32, 74)[tl]}
\put(1914,277){\oval(108,108)[bl]}
\put(1914,239){\oval( 74, 32)[br]}
}%
{\put(2439,239){\oval( 76, 12)[bl]}
\put(2439,314){\oval(224,162)[br]}
}%
{\put(3151,-399){\oval( 12, 76)[tl]}
\put(3226,-399){\oval(162,224)[bl]}
}%
{\put(1839,-361){\oval( 76, 12)[bl]}
\put(1839,-286){\oval(224,162)[br]}
}%
{\put(2664,-286){\oval(226,162)[bl]}
\put(2664,-361){\oval( 74, 12)[br]}
}%
{\put(2251,464){\circle{150}}
}%
{\put(976,-136){\circle{150}}
}%
{\put(1726,-136){\circle{150}}
}%
{\put(2776,-136){\circle{150}}
}%
{\put(3526,-136){\circle{150}}
}%
{\put(2251,-736){\circle{150}}
}%
{\put(2251,-1486){\circle{150}}
}%
{\put(2176,464){\vector(-2,-1){1170}}
}%
{\put(2251,389){\vector( 1,-1){450}}
}%
{\put(2326,464){\vector( 2,-1){1110}}
}%
{\put(976,-211){\vector( 1,-1){1200}}
}%
{\put(1051,-136){\vector( 2,-1){1140}}
}%
{\put(1726,-211){\vector( 1,-2){585}}
}%
{\put(1801,-136){\vector( 3,-4){414}}
}%
{\put(3451,-211){\vector(-2,-1){1110}}
}%
{\put(2176,389){\vector(-1,-1){450}}
}%
{\put(2701,-136){\vector(-3,-4){414}}
}%
{\put(2776,-211){\vector(-1,-3){405}}
}%
{\put(3526,-211){\vector(-1,-1){1237.500}}
}%
\put(2176,614){\makebox(0,0)[lb]{\smash{\SetFigFont{12}{14.4}
{\rmdefault}{\mddefault}{\updefault}{$A$}%
}}}
\put(2176,-1786){\makebox(0,0)[lb]{\smash{\SetFigFont{12}{14.4}
{\rmdefault}{\mddefault}{\updefault}{$B$}%
}}}
\put(751,-211){\makebox(0,0)[lb]{\smash{\SetFigFont{12}{14.4}
{\rmdefault}{\mddefault}{\updefault}{$v_1$}%
}}}
\put(2926,-211){\makebox(0,0)[lb]{\smash{\SetFigFont{12}{14.4}
{\rmdefault}{\mddefault}{\updefault}{$u_1$}%
}}}
\put(3676,-211){\makebox(0,0)[lb]{\smash{\SetFigFont{12}{14.4}
{\rmdefault}{\mddefault}{\updefault}{$u_2$}%
}}}
\put(1876,-211){\makebox(0,0)[lb]{\smash{\SetFigFont{12}{14.4}
{\rmdefault}{\mddefault}{\updefault}{$v_2$}%
}}}
\put(2251,-961){\makebox(0,0)[lb]{\smash{\SetFigFont{12}{14.4}
{\rmdefault}{\mddefault}{\updefault}{$v$}%
}}}
\end{picture}
\end{center}
\caption{The hypergraph $H(V,E_h)$ in Example \ref{reliablebutnotprivate}}
\label{figure5}
\end{figure}

Following Franklin and Yung \cite{fy}, and, Franklin and Wright \cite{fw},
we consider multicast as our only communication
primitive in this section. A message that is multicast by any node $A$
in a hypergraph is received by all nodes $A^*$  with privacy
(that is, nodes not in $A^*$ learn nothing about what was sent)
and authentication (that is, nodes in $A^*$ are guaranteed
to receive the value that was multicast and to know
which node multicast it).
We assume that all nodes in the
hypergraph know the complete protocol
specification and the complete structure of the hypergraph.
\begin{definition}\label{definition-main}
Let $H(V,E)$ be a hypergraph, $A,B \in V$ be distinct nodes of $H$, and
$k\geq 0$. $A$, $B$ are $k$-{\em separable\/} in $H$
if there is a node set $W\subset V$ with at most $k$ nodes
such that any directed path from $A$ to $B$ goes
through at least one node in $W$. We say that $W$ {\em separates\/} $A,B$.
\end{definition}
{\bf Remark.} Note that there is no straightforward relationship between
strong connectivity and separability in hypergraphs.
\begin{theorem}
\label{mainlemma}
Let $H(V,E)$ be a hypergraph, $A,B \in V$ be distinct nodes of $H$, and
$k\geq 0$. 
The nodes $A$ and $B$ are not $2k$-separable
if and only if there are $2k+1$ directed node disjoint
paths from $A$ to $B$ in $H$.
\end{theorem}
\begin{proof} This follows directly from the maximum-flow
minimum-cut theorem in classical graph theory. For details,
see, e.g., \cite{ff}. \hfill \mbox{Q.E.D.}\end{proof}
\begin{theorem}
\label{iffcondition}
Let $H(V,E)$ be a hypergraph, $A,B \in V$ be distinct nodes of $H$, and
$k\geq 0$. 
A necessary and sufficient condition for reliable message 
transmission
from $A$ to $B$ against a $k$-active adversary is that $A$ and $B$
are not $2k$-separable in $H$.
\end{theorem}\begin{proof}
First assume that $A$ and $B$ cannot be separated by a $2k$-node set.
By Theorem \ref{mainlemma}, there are  $2k+1$ directed node disjoint
paths from $A$ to $B$ in $H$. Thus reliable message transmission from
$A$ to $B$ is possible.

Next assume that $A$ and $B$ can be separated  by a $2k$-node set $W$ in $H$.
We shall show that reliable message transmission
is impossible.
Suppose that $\pi$ is a message
transmission protocol from $A$ to $B$ and
let $W=W_0\cup W_1$ be a
$2k$-node separation of $A$ and $B$ with  $W_0$ and  $W_1$ each
having at most $k$ nodes. Let $m_0$ be the message that $A$ transmits.
The adversary will attempt to maintain a simulation
of the possible behavior of $A$ by executing
$\pi$ for message $m_1\not= m_0$.
The strategy of the adversary is to flip a coin and then,
depending on the outcome, decide which set of $W_0$ or $W_1$ to control.
Let $W_b$ be the chosen set. In each execution  step of the
transmission protocol, the adversary causes each node in
$W_b$ to follow the protocol
$\pi$ as if the protocol were transmitting the message $m_1$.
This simulation succeeds with nonzero probability.
Since $B$ does not know whether $b=0$ or $b=1$,
at the end of the protocol $B$ cannot decide whether
$A$ has transmitted $m_0$ or $m_1$ if the adversary succeeds.
Thus with nonzero probability, the reliability
is not achieved. \hfill{Q.E.D.}\end{proof}
Theorem \ref{iffcondition} gives a sufficient and necessary
condition for achieving reliable message transmission against a $k$-active
adversary over hypergraphs. In the following example, we show that
this condition is not sufficient for achieving privacy against a $k$-active
adversary (indeed, even not for a $k$-passive adversary).
\begin{ex}
\label{reliablebutnotprivate}
Let $H(V,E_h)$ be the hypergraph in Figure \ref{figure5}
where $V=\{A$, $B$, $v_1$, $v_2$, $v$, $u_1$, $u_2\}$ and
$E_h=\{(A,\{v_1,v_2\})$, $(v_1,\{v,B\})$, $(v_2,\{v,B\})$,
$(A,\{u_1,u_2\})$, $(u_1,\{v,B\})$, $(u_2,\{v,B\})\}$.
Then the nodes $A$ and $B$ are not 2-separable in $H$.
Theorem \ref{iffcondition} shows that reliable message transmission
from $A$ to $B$ is possible against a 1-active adversary.
However, the hypergraph $H$ is not weakly 2-connected (the removal
of the node $v$ and the removal of the corresponding
hyperedges will disconnect $A$ and $B$). Thus, the result by
Franklin and Yung \cite{fy} shows that private message transmission
from $A$ to $B$ is not possible against a 1-passive adversary.
\end{ex}

\begin{theorem}
\label{prohy}
Let $\delta> 0$ and $A$ and $B$ be two nodes in a
hypergraph $H(V,E)$ satisfying
the following conditions:
\begin{enumerate}
\item $A$ and $B$ are not $2k$-separable in $H$.
\item $B$ and $A$ are not $2k$-separable in $H$.
\item $A$ and $B$ are strongly $k$-connected in $H$.
\end{enumerate}
Then there is a $(0,\delta)$-secure message transmission protocol
from $A$ to $B$ against a $k$-active adversary.
\end{theorem}
\begin{proof} 
Assume that the conditions of the theorem is satisfied.
For each $k$-node subset set $S$ of $V\setminus \{A,B\}$, let
$p_S$ be a directed path from $A$ to $B$ which witnesses that
$A$ and $B$ are strongly $k$-connected by removing the nodes in $S$ and
corresponding hyperedges in $H$. Let
${\cal S}=\{S:  S\subset V\setminus \{A,B\},|S|=k \}$ and
${\cal P}=\{p_S: S\in{\cal S}\}$. Then $A$ transmits
the message $M^A$ to $B$ using the following protocol.
\begin{algorithm}{9}
\item For each $S\in {\cal S}$, $A$
chooses a random pair $(a_S,b_S)\in_R {\bf F}^2$, and transmits
this pair to $B$ via the path $p_S$.
\item For each $S\in {\cal S}$,
$B$ receives a pair $(a_S^B,b_S^B)$ from $A$ via the path $p_S$.
\item  For each $S\in {\cal S}$, $B$ chooses a random
$r_S\in_R {\bf F}$ and computes $s_S=\auth(r_S; a_S^B,b_S^B)$.
\item $B$ reliably transmits $s=\langle \langle r_S,s_S\rangle:
S\in{\cal S} \rangle$ to $A$.
\item $A$ reliably receives the value
$s=\langle \langle r_S, s_S\rangle: S\in{\cal S} \rangle$ from $B$.
\item  $A$ computes the key index set $K_{\mathrm{index}}
=\{i_S: s_S=\auth(r_S;a_S^A,b_S^A)\}$ and
the shared secret $K^A=\sum_{i_S\in K_{\mathrm{index}}}a^A_S$.
\item $A$ reliably transmits $\langle K_{\mathrm{index}}, M^A+K^A\rangle$
to $B$, where $M^A$ is the secret message.
\item $B$ reliably receives the value 
$\langle K_{\mathrm{index}}, c\rangle$ from $A$.
$B$ computes the shared secret $K^B=\sum_{i_S\in K_{\mathrm{index}}}a^B_S$,
and decrypts the message $M^B=c-K^B$.
\end{algorithm}
It is possible that  $a^A_S\not= a^B_S$ but
$\auth(r_S; a^A_S,b^A_S)=\auth(r_S; a^B_S,b^B_S)$
for some $S\in{\cal S}$. However this probability is negligible.
Thus the above protocol is reliable with high probability.
Since $A$ and $B$ are strongly $k$-connected in $H$, there
is a pair $(a_S,b_S)$ such that  $(a_S,b_S)$ reliably reaches
$B$ and the adversary cannot infer any information of $a_S$ from
its view. Thus the above protocol is $(0,\delta)$-secure against
a $k$-active adversary if one chooses sufficiently large {\bf F}.
\hfill{Q.E.D.}\end{proof}
The results in Sections \ref{digraphsecd} and \ref{digraphsec}
show that the condition in Theorem \ref{prohy} is not necessary.
\section{Secure message transmission over neighbor networks}
\label{neighborsec}
\subsection{Definitions}
A special case of the hypergraph is the {\em neighbor networks}.
A neighbor network is a graph $G(V,E)$. In a neighbor network,
a node $A\in V$ is called a neighbor of another node $B\in V$
if there is an edge $(A,B)\in E$.
In a neighbor network, we assume that
any message sent by a node $A$ will be received
identically by all its neighbors, whether or not
$A$ is faulty, and all parties outside
of $A$'s neighbor learn nothing about the content of
the message.

For a neighbor network $G(V,E)$ and two nodes $A,B$ in it,
Franklin and Wright \cite{fw}, and, Wang and Desmedt \cite{wd}
showed that if there are $n$ multicast lines (that is, $n$ paths
with disjoint neighborhoods)
between $A$ and $B$ and there
are at most $k$ malicious (Byzantine style) processors,
then the condition $n>k$ is necessary and sufficient
for achieving efficient probabilistically reliable
and perfect private communication.

For each neighbor network $G(V,E)$, there is a hypergraph
$H_G(V,E_h)$ which is equivalent to $G(V,E)$ in functionality.
$H_G(V,E_h)$ is defined by letting $E_h$ be the set of
hyperedges $(A,A^*)$ where $A\in V$ and  $A^*$ is the set of
neighbors of $A$.

\begin{figure}[htb]
\begin{center}
\setlength{\unitlength}{3947sp}%
\begingroup\makeatletter\ifx\SetFigFont\undefined%
\gdef\SetFigFont#1#2#3#4#5{%
  \reset@font\fontsize{#1}{#2pt}%
  \fontfamily{#3}\fontseries{#4}\fontshape{#5}%
  \selectfont}%
\fi\endgroup%
\begin{picture}(1650,1035)(451,-361)
{\thinlines
\put(676,164){\circle{150}}
}%
{\put(1351,464){\circle{150}}
}%
{\put(1351,-136){\circle{150}}
}%
{\put(1951,164){\circle{150}}
}%
{\put(751,239){\line( 5, 2){530.172}}
}%
{\put(751, 89){\line( 5,-2){530.172}}
}%
{\put(1426,-136){\line( 2, 1){450}}
}%
{\put(1426,464){\line( 5,-2){530.172}}
}%
{\put(1351,389){\line( 0,-1){450}}
}%
\put(451, 89){\makebox(0,0)[lb]{\smash{\SetFigFont{12}{14.4}
{\rmdefault}{\mddefault}{\updefault}{$A$}%
}}}
\put(2101, 89){\makebox(0,0)[lb]{\smash{\SetFigFont{12}{14.4}
{\rmdefault}{\mddefault}{\updefault}{$B$}%
}}}
\put(1276,-361){\makebox(0,0)[lb]{\smash{\SetFigFont{12}{14.4}
{\rmdefault}{\mddefault}{\updefault}{$C$}%
}}}
\put(1276,539){\makebox(0,0)[lb]{\smash{\SetFigFont{12}{14.4}
{\rmdefault}{\mddefault}{\updefault}{$D$}%
}}}
\end{picture}
\end{center}
\caption{The graph $G(V,E)$ in Example \ref{exforcon}}
\label{figure1}
\end{figure}
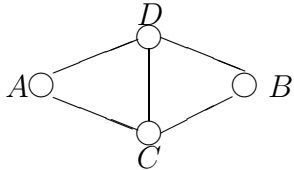

Let $A$ and $B$ be two nodes in a neighbor network $G(V,E)$.
We have the following definitions:
\begin{enumerate}
\item $A$ and $B$ are $k$-{\em connected} in $G(V,E)$ if there are
$k$ node disjoint paths between $A$ and $B$ in $G(V,E)$.
\item $A$ and $B$ are {\em weakly $k$-hyper-connected} in  $G(V,E)$ if
$A$ and $B$ are weakly $k$-connected in $H_G(V,E_h)$.
\item  $A$ and $B$ are $k$-{\em neighbor-connected} in $G(V,E)$ if for any
set $V_1\subseteq V\setminus\{A,B\}$ with $|V_1|<k$, the removal
of $neighbor(V_1)$ and all incident edges
{from} $G(V,E)$ does not disconnect $A$ and $B$, where
$$neighbor(V_1)=V_1\cup\{A\in V:
\mathrm{ there exists} B\in V_1 (B,A)\mbox{ such that }\in E\}
\setminus\{A,B\}.$$
\item $A$ and $B$ are {\it weakly $(n,k)$-connected} if there are
$n$ node disjoint paths $p_1,\allowbreak  \ldots,\allowbreak  p_n$
between $A$ and $B$ and, for any node set
$T\subseteq (V\setminus \{A,B\})$
with $|T|\le k$, there exists $1\le i\le n$
such that all nodes on $p_i$ have no neighbor in $T$.
\end{enumerate}
It is easy to check that the following relationships hold.
\begin{center}
weak $(n,k-1)$-connectivity $(n\ge k)$
$\Rightarrow$ $k$-neighbor-connectivity
$\Rightarrow$ weak $k$-hyper-connectivity
$\Rightarrow$ $k$-connectivity
\end{center}
In the following examples, we show that these implications
are strict.
\begin{ex}
\label{exforcon}
Let $G(V,E)$ be the graph in Figure \ref{figure1} where
$V=\{A, B, C,D\}$ and $E=\{(A,C), (C,B),(A,D),(D,B), (C,D)\}$.
Then it is straightforward to check that $G(V,E)$ is 2-connected
but not weakly 2-hyper-connected.
\end{ex}

\begin{figure}[htb]
\begin{center}
\setlength{\unitlength}{3947sp}%
\begingroup\makeatletter\ifx\SetFigFont\undefined%
\gdef\SetFigFont#1#2#3#4#5{%
  \reset@font\fontsize{#1}{#2pt}%
  \fontfamily{#3}\fontseries{#4}\fontshape{#5}%
  \selectfont}%
\fi\endgroup%
\begin{picture}(1500,1185)(226,-436)
{\thinlines
\put(451,239){\circle{150}}
}%
{\put(1051,539){\circle{150}}
}%
{\put(1051,239){\circle{150}}
}%
{\put(1051,-136){\circle{150}}
}%
{\put(1576,239){\circle{150}}
}%
{\put(526,314){\line( 2, 1){450}}
}%
{\put(1126,-136){\line( 3, 2){450}}
}%
{\put(526,164){\line( 4,-3){516}}
}%
{\put(1126,539){\line( 2,-1){450}}
}%
{\put(1051,464){\line( 0,-1){150}}
}%
{\put(1051,164){\line( 0,-1){225}}
}%
\put(226,164){\makebox(0,0)[lb]{\smash{\SetFigFont{12}{14.4}
{\rmdefault}{\mddefault}{\updefault}{$A$}%
}}}
\put(1726,164){\makebox(0,0)[lb]{\smash{\SetFigFont{12}{14.4}
{\rmdefault}{\mddefault}{\updefault}{$B$}%
}}}
\put(976,-436){\makebox(0,0)[lb]{\smash{\SetFigFont{12}{14.4}
{\rmdefault}{\mddefault}{\updefault}{$C$}%
}}}
\put(976,614){\makebox(0,0)[lb]{\smash{\SetFigFont{12}{14.4}
{\rmdefault}{\mddefault}{\updefault}{$D$}%
}}}
\put(901, 14){\makebox(0,0)[lb]{\smash{\SetFigFont{12}{14.4}
{\rmdefault}{\mddefault}{\updefault}{$F$}%
}}}
\end{picture}
\end{center}
\caption{The graph $G(V,E)$ in Example \ref{exforwd}}
\label{figure2}
\end{figure}
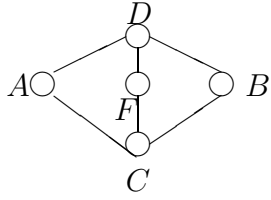

\begin{ex}
\label{exforwd}
Let $G(V,E)$ be the graph in Figure \ref{figure2} where $V =\{A, B, C,D,F\}$
and $E =\{(A,C)$, $(A,D)$, $(C,B)$, $(D,B)$, $(C,F)$, $(F,D)\}$.
Then it is straightforward to check that $A$ and $B$ are weakly
$2$-hyper-connected but not $2$-neighbor-connected.
\end{ex}

\begin{figure}[htb]
\begin{center}
\setlength{\unitlength}{3947sp}%
\begingroup\makeatletter\ifx\SetFigFont\undefined%
\gdef\SetFigFont#1#2#3#4#5{%
  \reset@font\fontsize{#1}{#2pt}%
  \fontfamily{#3}\fontseries{#4}\fontshape{#5}%
  \selectfont}%
\fi\endgroup%
\begin{picture}(2775,1110)(151,-436)
{\thinlines
\put(376,164){\circle{150}}}%
{\put(976,464){\circle{150}}}%
{\put(1576,464){\circle{150}}}%
{\put(2176,464){\circle{150}}}%
{\put(976,-136){\circle{150}}}%
{\put(1576,-136){\circle{150}}}%
{\put(2176,-136){\circle{150}}}%
{\put(2776,164){\circle{150}}}%
{\put(451,239){\line( 2, 1){450}}}%
{\put(1051,464){\line( 1, 0){450}}}
{\put(1651,464){\line( 1, 0){450}}}%
{\put(2251,464){\line( 2,-1){450}}}%
{\put(451, 89){\line( 2,-1){450}}}%
{\put(1051,-136){\line( 1, 0){450}}}%
{\put(1651,-136){\line( 1, 0){450}}}%
{\put(2251,-136){\line( 5, 2){530.172}}}%
{\put(1051,464){\line( 2, -1){1080}}}
{\put(1051,-136){\line( 2, 1){1080}}}
\put(151, 89){\makebox(0,0)[lb]{\smash{\SetFigFont{12}{14.4}
{\rmdefault}{\mddefault}{\updefault}{$A$}%
}}}
\put(2926, 89){\makebox(0,0)[lb]{\smash{\SetFigFont{12}{14.4}
{\rmdefault}{\mddefault}{\updefault}{$B$}%
}}}
\put(901,539){\makebox(0,0)[lb]{\smash{\SetFigFont{12}{14.4}
{\rmdefault}{\mddefault}{\updefault}{$C$}%
}}}
\put(1501,539){\makebox(0,0)[lb]{\smash{\SetFigFont{12}{14.4}
{\rmdefault}{\mddefault}{\updefault}{$D$}%
}}}
\put(2101,539){\makebox(0,0)[lb]{\smash{\SetFigFont{12}{14.4}
{\rmdefault}{\mddefault}{\updefault}{$E$}%
}}}
\put(901,-436){\makebox(0,0)[lb]{\smash{\SetFigFont{12}{14.4}
{\rmdefault}{\mddefault}{\updefault}{$F$}%
}}}
\put(1501,-436){\makebox(0,0)[lb]{\smash{\SetFigFont{12}{14.4}
{\rmdefault}{\mddefault}{\updefault}{$G$}%
}}}
\put(2101,-436){\makebox(0,0)[lb]{\smash{\SetFigFont{12}{14.4}
{\rmdefault}{\mddefault}{\updefault}{$H$}%
}}}
\end{picture}

\end{center}
\caption{The graph $G(V,E)$ in Example \ref{80ex}}
\label{figure80}
\end{figure}
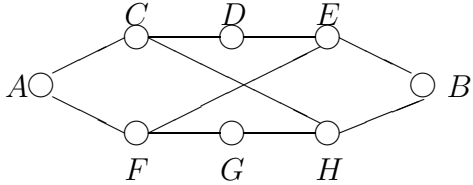

\begin{ex}
\label{80ex}
Let $G(V,E)$ be the graph in Figure \ref{figure80} where $V =\{A$,  $B$,
$C$, $D$, $E$, $F$, $G$, $H\}$
and $E =\{(A,C)$, $(C,D)$, $(D,E)$ $(E,B)$,
$(A,F)$, $(F,G)$, $(G,H)$ $(H,B)$,
$(C,H)$, $(E,F)\}$.
Then it is straightforward to check that $A$ and $B$ are
$2$-neighbor-connected but not weakly $(2,1)$-connected.
\end{ex}

Example \ref{exforcon} shows that $k$-connectivity does
not necessarily imply weak $k$-hyper-connectivity.
Example \ref{exforwd} shows that weak  $k$-hyper-connectivity
does not necessarily imply $k$-neighbor-connectivity.
Example \ref{80ex} shows that $k$-neighbor connectivity
does not necessarily imply weak $(n,k-1)$-connectivity for some $n\ge k$.
\subsection{$(0,\delta)$-Secure message transmission over neighbor networks}
Wang and Desmedt \cite{wd} have given a sufficient condition
for achieving $(0,\delta)$-security message transmission
against a $k$-active adversary over neighbor networks. In this section,
we show that their condition is not necessary.
\begin{theorem}
\label{ntwk}
(Wang and Desmedt \cite{wd})
If $A$ and $B$ are  {\it weakly $(n,k)$-connected}
for some $k<n$, then there is an efficient $(0,\delta)$-secure
message transmission between $A$ and $B$.
\end{theorem}

The condition in Theorem \ref{ntwk} is not necessary. For example,
the neighbor network $G$ in Example \ref{exforwd} is not
$2$-neighbor-connected, thus not weakly $(2,1)$-connected. In the following
we present a $(0,\delta)$-secure message transmission  protocol 
against a 1-active adversary from $A$ to $B$ for the neighbor network 
of Example \ref{exforwd} .

\vskip 5pt
\noindent
{\bf Message transmission protocol for neighbor network $G$ in
Example \ref{exforwd}.}

\begin{algorithm}{99}
\item $A$ chooses two random pairs $(r^A_1, r^A_2)\in_R {\bf F}^2$ and
$(r^A_3, r^A_4)\in_R {\bf F}^2$. $A$ sends  $(r^A_1, r^A_2)$ to $C$ and
$(r^A_3, r^A_4)$ to $D$.
\item $B$ chooses two random pairs $(r^B_1, r^B_2)\in_R {\bf F}^2$ and
$(r^B_3, r^B_4)\in_R {\bf F}^2$. $B$ sends  $(r^B_1, r^B_2)$ to $C$ and
$(r^B_3, r^B_4)$ to $D$.
\item $C$ chooses a random pair $(a_1,b_1)\in_R {\bf F}^2$. $C$ sends
$(a_1+r^A_1, b_1+r^A_2)$ to $A$ and $(a_1+r^B_1, b_1+r^B_2)$ to $B$.
\item  $D$ chooses a random pair $(a_2,b_2)\in_R {\bf F}^2$. $D$ sends
$(a_2+r^A_3, b_2+r^A_4)$ to $A$ and $(a_2+r^B_3, b_2+r^B_4)$ to $B$.
\item From the messages received from $C$ and $D$, $A$ computes
$(a_1^A,b_1^A)$ and $(a_2^A,b_2^A)$.
\item From the messages received from $C$ and $D$, $B$ computes
$(a_1^B,b_1^B)$ and $(a_2^B,b_2^B)$.
\item $B$ chooses a random $r\in_R {\bf F}$,
computes $s_1=\auth(r; a^B_1,b^B_1)$
and  $s_2=\auth(r; a^B_2,b^B_2)$. Using the probabilistically
reliable message transmission protocol of Franklin and Wright \cite{fw},
$B$ transmits $\langle r, s_1, s_2\rangle$ to $A$.
\item Let  $\langle r^A, s^A_1, s^A_2\rangle$ be the message
received by $A$ in the last step, $A$ computes
the key index set $K_{\mathrm{index}}
=\{i: s^A_i=\auth(r^A;a_i^A,b_i^A)\}$.
$A$ also computes the shared secret $K^A=\sum_{i\in
K_{\mathrm{index}}}a^A_i$.
\item Using the probabilistically reliable message transmission protocol
of Franklin and Wright \cite{fw}, $A$
transmits $\langle K_{\mathrm{index}}, M^A+K^A\rangle$ to $B$,
where $M^A$ is the secret message.
\item Let $\langle K^B_{\mathrm{index}}, c^B\rangle$
be the message that $B$ received in the last step. $B$ computes the shared
secret $K^B=\sum_{i\in K^B_{\mathrm{index}}}a^B_i$,
and decrypts the message $M^B=c^B-K^B$.
\end{algorithm}

It is straightforward to check that the above protocol is an efficient
$(0,\delta)$-secure message transmission protocol from $A$ to $B$
against a 1-active adversary.

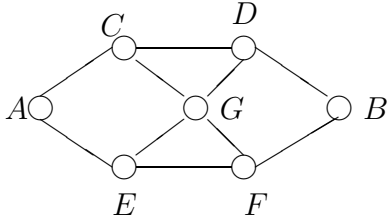
\begin{figure}[htb]
\begin{center}
\setlength{\unitlength}{3947sp}%
\begingroup\makeatletter\ifx\SetFigFont\undefined%
\gdef\SetFigFont#1#2#3#4#5{%
  \reset@font\fontsize{#1}{#2pt}%
  \fontfamily{#3}\fontseries{#4}\fontshape{#5}%
  \selectfont}%
\fi\endgroup%
\begin{picture}(2250,1335)(76,-586)
{\thinlines
\put(301, 89){\circle{150}}
}%
{\put(826,464){\circle{150}}
}%
{\put(1576,464){\circle{150}}
}%
{\put(2176, 89){\circle{150}}
}%
{\put(826,-286){\circle{150}}
}%
{\put(1576,-286){\circle{150}}
}%
{\put(1276, 89){\circle{150}}
}%
{\put(301,164){\line( 3, 2){450}}
}%
{\put(301, 14){\line( 3,-2){450}}
}%
{\put(901,-286){\line( 1, 0){600}}
}%
{\put(901,464){\line( 1, 0){600}}
}%
{\put(1651,464){\line( 3,-2){450}}
}%
{\put(1651,-286){\line( 5, 3){518.382}}
}%
{\put(901,-211){\line( 4, 3){300}}
}%
{\put(1351,164){\line( 1, 1){225}}
}%
{\put(901,389){\line( 4,-3){300}}
}%
{\put(1351, 14){\line( 1,-1){225}}
}%
\put( 76, 14){\makebox(0,0)[lb]{\smash{\SetFigFont{12}{14.4}
{\rmdefault}{\mddefault}{\updefault}{$A$}%
}}}
\put(2326, 14){\makebox(0,0)[lb]{\smash{\SetFigFont{12}{14.4}
{\rmdefault}{\mddefault}{\updefault}{$B$}%
}}}
\put(676,539){\makebox(0,0)[lb]{\smash{\SetFigFont{12}{14.4}
{\rmdefault}{\mddefault}{\updefault}{$C$}%
}}}
\put(1501,614){\makebox(0,0)[lb]{\smash{\SetFigFont{12}{14.4}
{\rmdefault}{\mddefault}{\updefault}{$D$}%
}}}
\put(751,-586){\makebox(0,0)[lb]{\smash{\SetFigFont{12}{14.4}
{\rmdefault}{\mddefault}{\updefault}{$E$}%
}}}
\put(1576,-586){\makebox(0,0)[lb]{\smash{\SetFigFont{12}{14.4}
{\rmdefault}{\mddefault}{\updefault}{$F$}%
}}}
\put(1426, 14){\makebox(0,0)[lb]{\smash{\SetFigFont{12}{14.4}
{\rmdefault}{\mddefault}{\updefault}{$G$}%
}}}
\end{picture}

\end{center}
\caption{The graph $G(V,E)$ in Example \ref{ness}}
\label{figure3}
\end{figure}

Example \ref{reliablebutnotprivate} shows that for a general
hypergraph, the existence of a reliable message transmission
protocol does not imply the existence of a private message transmission
protocol. We show that this is true for probabilistic reliability
and perfect privacy in neighbor networks also.
\begin{ex}
\label{ness}
Let $G(V,E)$ be the neighbor network in Figure \ref{figure3} where
$V=\{A,\allowbreak B,\allowbreak C,\allowbreak D,\allowbreak E,\allowbreak
F,\allowbreak G\}$ and
$E=\{(A,C),\allowbreak (C,D),\allowbreak  (D,B),\allowbreak
(A,E),\allowbreak  (E,F),\allowbreak  (F,B),\allowbreak
(G,C),\allowbreak (G,D),\allowbreak (G,E),\allowbreak (G,F)\}$.
Then there is a probabilistic reliable message transmission
protocol from $A$ to $B$ against a 1-active adversary in $G$. But there is
no
private message transmission from $A$ to $B$ against a 1-passive (or
1-active)
adversary in $G$.
\end{ex}

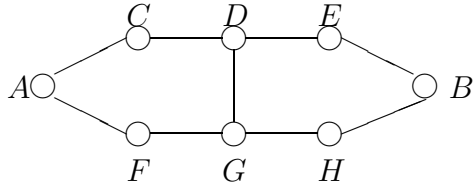
\begin{figure}[htb]
\begin{center}
\setlength{\unitlength}{3947sp}%
\begingroup\makeatletter\ifx\SetFigFont\undefined%
\gdef\SetFigFont#1#2#3#4#5{%
  \reset@font\fontsize{#1}{#2pt}%
  \fontfamily{#3}\fontseries{#4}\fontshape{#5}%
  \selectfont}%
\fi\endgroup%
\begin{picture}(2775,1110)(151,-436)
{\thinlines
\put(376,164){\circle{150}}}%
{\put(976,464){\circle{150}}}%
{\put(1576,464){\circle{150}}}%
{\put(2176,464){\circle{150}}}%
{\put(976,-136){\circle{150}}}%
{\put(1576,-136){\circle{150}}}%
{\put(2176,-136){\circle{150}}}%
{\put(2776,164){\circle{150}}}%
{\put(451,239){\line( 2, 1){450}}}%
{\put(1051,464){\line( 1, 0){450}}}
{\put(1651,464){\line( 1, 0){450}}}%
{\put(2251,464){\line( 2,-1){450}}}%
{\put(451, 89){\line( 2,-1){450}}}%
{\put(1051,-136){\line( 1, 0){450}}}%
{\put(1651,-136){\line( 1, 0){450}}}%
{\put(2251,-136){\line( 5, 2){530.172}}}%
{\put(1576,389){\line( 0,-1){450}}}%
\put(151, 89){\makebox(0,0)[lb]{\smash{\SetFigFont{12}{14.4}
{\rmdefault}{\mddefault}{\updefault}{$A$}%
}}}
\put(2926, 89){\makebox(0,0)[lb]{\smash{\SetFigFont{12}{14.4}
{\rmdefault}{\mddefault}{\updefault}{$B$}%
}}}
\put(901,539){\makebox(0,0)[lb]{\smash{\SetFigFont{12}{14.4}
{\rmdefault}{\mddefault}{\updefault}{$C$}%
}}}
\put(1501,539){\makebox(0,0)[lb]{\smash{\SetFigFont{12}{14.4}
{\rmdefault}{\mddefault}{\updefault}{$D$}%
}}}
\put(2101,539){\makebox(0,0)[lb]{\smash{\SetFigFont{12}{14.4}
{\rmdefault}{\mddefault}{\updefault}{$E$}%
}}}
\put(901,-436){\makebox(0,0)[lb]{\smash{\SetFigFont{12}{14.4}
{\rmdefault}{\mddefault}{\updefault}{$F$}%
}}}
\put(1501,-436){\makebox(0,0)[lb]{\smash{\SetFigFont{12}{14.4}
{\rmdefault}{\mddefault}{\updefault}{$G$}%
}}}
\put(2101,-436){\makebox(0,0)[lb]{\smash{\SetFigFont{12}{14.4}
{\rmdefault}{\mddefault}{\updefault}{$H$}%
}}}
\end{picture}

\end{center}
\caption{The graph $G(V,E)$}
\label{figure009}
\end{figure}

\begin{proof}
It is straightforward to check that $G(V,E)$ is not weakly
2-hyper-connected.
Indeed, in the hypergraph $H_G(V,E_h)$ of $G(V,E)$, the removal of node
$G$ and the removal of the corresponding hyperedges
will disconnect $A$ and $B$ completely. Thus Franklin and
Yung's result in \cite{fy} shows that there is no private message
transmission
protocol against a 1-passive (or 1-active) adversary from $A$ to $B$.
It is also straightforward to check that Franklin and Wright's
\cite{fw} reliable message transmission protocol against a
1-active adversary works for the two paths $(A,C,D,B)$ and
$(A,E,F,B)$.
\hfill{Q.E.D.}\end{proof}
Though weak $k$-hyper-connectivity is a necessary
condition for achieving probabilistically reliable and perfectly
private message transmission against a $(k-1)$-active adversary, we do not
know whether this condition is sufficient. We conjecture that there
is no probabilistically reliable and perfectly
private message transmission protocol against a 1-active
adversary for the weakly 2-hyper-connected neighbor network $G(V,E)$ in
Figure \ref{figure009}, where $V=\{A$, $B$, $C$, $D$, $E$, $F$, $G$, $H\}$
and
$E=\{(A,C)$, $(C,D)$, $(D,E)$, $(E,B)$, $(A,F)$, $(F,G)$, $(G,H)$,
$(H,B)$, $(D,G)\}$. Note that in order to prove or refute our conjecture,
it is sufficient to show whether there is a probabilistically
reliable message transmission protocol against a 1-active
adversary for the neighbor network. For this specific neighbor network,
the trick in our previous protocol could be used to convert
any probabilistically reliable message transmission protocol
to a  probabilistically reliable and perfectly
private message transmission protocol against a 1-active
adversary.

\end{document}